\documentclass[preprint,12pt]{elsarticle}
\usepackage[utf8]{inputenc}
\usepackage[T1]{fontenc}
\usepackage{mathptmx}
\usepackage{amsmath,amssymb,amsfonts}
\usepackage{siunitx}
\usepackage{graphicx}
\graphicspath{{Figures/}}
\usepackage{booktabs}
\usepackage{multirow}
\usepackage{tabularx}
\usepackage{longtable}
\usepackage{caption}
\usepackage{subcaption}
\usepackage{geometry}
\usepackage{float}
\usepackage{comment}
\usepackage[section]{placeins}  
\usepackage[version=4]{mhchem}
\usepackage{xcolor}
\newcommand{\rev}[1]{#1}
\colorlet{red}{black} 
\definecolor{revred}{RGB}{255,0,0}
\newcommand{\revB}[1]{#1} 

\usepackage{hyperref}        
\usepackage[capitalise]{cleveref}  
\journal{Acta Materialia}
\biboptions{numbers,sort&compress}   
\begin{document}

\begin{frontmatter}

\title{Predicting Coherent B2 Stability in Ru-Containing Refractory Alloys Through Thermodynamic–Elastic Design Maps}

\author {Avik Mahata}
\ead{mahataa@merrimack.edu}

\cortext[cor1]{Corresponding author}

\address {Department of Mechanical and Electrical Engineering, Merrimack College, North Andover, MA, 01845, USA}

\begin{abstract}
The development of next-generation refractory superalloys is fundamentally constrained by competing requirements of high thermodynamic stability and low interfacial energy. Ru-based B2 intermetallics possess strong intrinsic ordering, but their use as coherent reinforcing phases requires balancing formation enthalpy against elastic strain. Here, we present a physics-guided machine-learning framework that integrates high-throughput density functional theory (DFT), Random Forest screening, and symbolic regression to map this design space. The approach addresses the ``binary paradox'' observed in recent experimental studies, where strongly ordered Ru-refractory compounds do not necessarily achieve the highest solvus temperatures. The resulting closed-form relation, $T_{\mathrm{solvus}} \approx 0.11(\Delta H/\Delta S)-20000|\delta|$, quantifies the competing effects of thermodynamic stabilization and lattice misfit. Because the electronic-structure calculations are static 0~K calculations, the relation is treated as an effective calibrated design equation rather than a directly calculated finite-temperature free-energy boundary. The model is benchmarked against published B2 solvus measurements, while calculated formation enthalpies and lattice parameters are independently compared with calorimetric and diffraction data. Within the fitted relation, a 1\% lattice misfit corresponds to an effective solvus penalty of approximately 200~$^\circ$C. Al/Si substitution further shows that improved lattice matching is accompanied by reduced thermodynamic stabilization, with zero misfit near 11~at.\% Al and 7~at.\% Si for Nb$_{75}$V$_{25}$. Higher-order refractory matrices and targeted electronic-structure calculations further support the transferability and physical interpretation of the framework. These results provide a quantitative route for identifying Ru-containing B2--BCC combinations that preserve strong chemical ordering while minimizing coherency penalties in refractory alloy design at high temperature.
\end{abstract}


\end{frontmatter}


\section{Introduction}

Refractory multi-principal element alloys that are based on group IV and group V elements such as Nb, Mo, Ta, Hf, Zr, and Ti are promising for use above about 1200~$^\circ$C, where Ni-based superalloys begin to approach intrinsic limits \cite{Jin2024,Ouyang2023,Senkov2019,Karimpilakkal2025,Frey2022a}. Their high melting points and complex phase stability make them attractive for aerospace, energy, and nuclear applications \cite{Senkov2018,Wang2023}. However, it is still difficult to achieve a combination of high-temperature strength, oxidation resistance, and acceptable room-temperature ductility \cite{Jin2024,Senkov2018,George2019_NatRevMat}. The intrinsic brittleness of many refractory BCC solid solutions remains a critical barrier to deployment, and identifying compositions that balance ductility with high-temperature retention requires navigating a vast compositional landscape \cite{Miracle2017_ActaMat, Sheikh2016_JAP}. Recent work has shown that coherent precipitation of ordered B2 phases in a disordered BCC matrix can provide a refractory analog to the classic $\gamma$ and $\gamma'$ architecture in Ni superalloys \cite{Yang2025,Soni2025,Sloof2019_AnnRev}. This A2 and B2 strategy is now a central design direction for refractory multi-principal element alloys.

Many early BCC and B2 superalloys used conventional B2 formers such as Al and Ti \cite{brodie2024stability, soni2025some, moczadlo2024microstructure, panina2021aging}. These elements can provide strong order hardening but often have relatively low B2 solvus temperatures and limited precipitate stability at very high temperatures \cite{Yang2025,Senkov2018}. Furthermore, excessive Al or Ti additions can promote the formation of brittle intermetallics like silicides or Laves phases which degrade fracture toughness \cite{Zhang2018_AdvEngMat}. Ruthenium-containing B2 compounds have recently emerged as a more robust choice. Experimental work has shown that HfRu and ZrRu B2 phases can form coherent precipitates within Nb and Nb-V based matrices and can remain stable at temperatures that exceed 1300~$^\circ$C and in some cases approach 1900~$^\circ$C \cite{Kube2024, FreyThesis,  Frey2022a}. These alloys show cuboidal or near cuboidal B2 precipitates with low lattice misfit and significant volume fraction over a broad processing window. They also demonstrate creep resistance that begins to rival advanced Ni-based systems \cite{Yang2025}. At the same time, in situ and post-mortem studies of BCC to B2 transformation pathways have revealed rich domain structures and variant selection that depend sensitively on temperature and composition \cite{Soni2025,Wang2024}.

Computational tools now play a central role in the design of these alloys. High-throughput and data-driven approaches link CALPHAD, first-principles calculations, and machine learning to explore large composition spaces \cite{Ouyang2023,Senkov2019,Feng2021,Kube2024}. DFT has been used to quantify the relative stability of B2 among refractory metals and to highlight the special role of group IV and late transition metal pairs such as Hf-Ru and Zr-Ru \cite{Wang2024}. Classical molecular dynamics has also been applied to related questions such as diffusion, elastic response, and defect behavior in complex alloys \cite{Plimpton1995}. At the mesoscale, coherent elastic misfit and its influence on phase separation, coarsening, and precipitate shape are now well established through continuum and phase-field models \cite{Fratzl1999,Jog2000}.

The application of machine learning (ML) has further accelerated this discovery process by overcoming the scaling limitations of traditional simulation methods. Recent studies have demonstrated the efficacy of ML in predicting phase formation, mechanical properties, and oxidation behavior in high-entropy alloys with accuracy comparable to expensive quantum mechanical calculations \cite{Wei2019_ActaMat, Wen2019_ActaMat, Islam2018_CompMatSci}. For example, Rao et al. successfully employed ML models to predict the yield strength of refractory multi-principal element alloys, while other works have utilized deep learning to classify phase stability boundaries in complex multi-component systems \cite{Rao2019_ActaMat, Zhou2019_npjCompMat, Kaufmann2020_ActaMat}. These data-centric approaches allow for the rapid screening of thousands of candidate compositions that would be computationally prohibitive to simulate directly \cite{Chen2020_JMST, Baty2019_Materials}. However, a recurring challenge in these "black-box" models is the lack of physical interpretability, which often obscures the underlying mechanistic drivers of stability such as lattice strain and chemical ordering \cite{Rickman2019_Nature, Schmidt2019_MRS}. Together, these efforts show that any predictive model for B2-strengthened refractory alloys must account for thermodynamic driving forces, elastic misfit, and microstructural morphology while maintaining interpretability.

Despite this progress, there are important gaps that limit the practical deployment of Ru-strengthened alloys. Ruthenium is scarce, dense, and expensive \cite{Ouyang2023,Kube2024}. Existing high-stability Ru-B2 alloys often require large Ru contents and therefore impose significant cost and density penalties \cite{Frey2022a,Frey2024a,Frey2024b}. At the same time, engineering alloys frequently include elements such as Al, Cr, Cu, and Si to improve oxidation resistance, adjust density, or enhance room-temperature ductility \cite{Wang2023,Feng2021,Jin2024}. The effects of these substitutions on Ru-based B2 solvus temperature, coherency, and morphology are only partially understood. Experiments on selected Nb-based Ru-B2 systems show that unwanted grain boundary phases and large lattice misfit can destroy coherency or promote cracking during processing \cite{Kube2024,Frey2025,FreyThesis}. However, there is no compact quantitative map that links specific chemical substitutions to changes in B2 stability and misfit.

In this work, we address this gap by building a physics-informed and data-efficient model for Ru-containing BCC and B2 alloys. We combine first-principles calculations of Ru-based B2 compounds with composition-derived descriptors and misfit-based criteria to construct a symbolic regression model for the B2 solvus temperature \cite{Ouyang2023,Feng2021,Wang2024}. The model is constrained by classical theory of coherent elastic misfit and precipitate shape transitions, following the ideas of Fratzl, Jog, and co-workers \cite{Fratzl1999,Jog2000}. \rev{We benchmark the model against published observations of HfRu- and ZrRu-containing BCC--B2 alloys in Nb and Nb--V matrices \cite{Frey2024b,Frey2024a,Frey2025,Kube2024}; these comparisons are external literature benchmarks and are not experiments performed in this study.} The resulting closed-form expression quantifies the stability penalty associated with lattice mismatch and allows for the identification of a design window where B2 precipitates remain coherent and thermally robust. We specifically investigate the impact of partial Ru substitution by Al, Cr, Cu, or Si to determine if high-temperature stability can be maintained in Ru-lean compositions. This framework provides a set of design rules that connect alloy chemistry, misfit, and precipitate morphology for the development of next-generation refractory superalloys. \rev{The dataset spans binary and equiatomic ternary BCC matrices, five equiatomic quaternary BCC SQS matrices, and the equiatomic quinary NbMoTaVW matrix, together with binary and multicomponent Ru-containing B2 precipitate chemistries. The higher-order BCC matrices are paired with the same B2 dataset to examine whether the lattice-matching design principle remains applicable as the refractory matrix becomes compositionally more complex.}

\section{Computational methodology}

\subsection{Density functional theory calculations}

\begin{figure}[H]
    \centering
    \includegraphics[width=\textwidth]{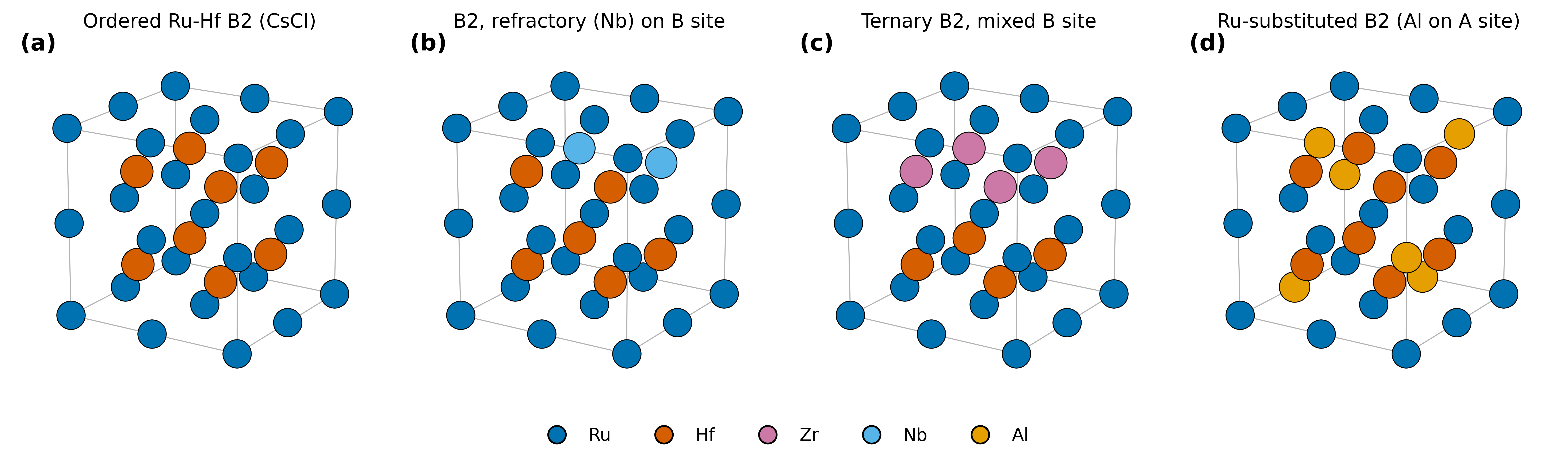}
\caption{\rev{Representative structural schematics for the B2 configurations used in this work. (a) is an ordered Ru--Hf B2 (CsCl-type) binary, (b) is a B2 cell with partial substitution of a refractory element on the B sublattice, (c) is a ternary B2 with mixed B-sublattice occupancy, and (d) is a Ru-substituted B2 with an Al dopant on the A sublattice. In panel (b), the minority species on the B sublattice is the refractory matrix element (here Nb). Ru occupies the A sublattice in all B2-based models. Corresponding real-space renderings of representative relaxed DFT supercells, with the complete element colour legend, are provided in Supplementary Fig.~S1.}}
    \label{fig:dft_structures}
\end{figure}

All first-principles calculations were performed using the Vienna \emph{Ab initio} Simulation
Package (VASP) \cite{ hafner2008ab} within the framework of density functional theory. Projector augmented wave
(PAW) potentials were employed together with the Perdew--Burke--Ernzerhof (PBE) \cite{ernzerhof1999assessment} generalized
gradient approximation for exchange and correlation (See Supplementary Material Tables S1 and S2 for more details). The overall DFT workflow consisted of
three stages: (i) systematic generation of BCC and B2 structural models, (ii) full relaxation of
cell shape and atomic positions, and (iii) static single-point total energy calculations on the
relaxed geometries. All calculations were organized in a hierarchical directory structure under
DFT Runs to ensure traceability and reproducibility. Baseline BCC reference cells were generated for all pure elements appearing in the alloy design
space, including Al, Cr, Cu, Hf, Mo, Nb, Ru, Si, Ta, Ti, V, W, and Zr. Each baseline calculation
used a conventional BCC unit cell with an initial lattice parameter of approximately 3.1~\AA,
which was subsequently fully relaxed. These elemental BCC structures serve as energetic
reference states for formation-energy calculations and also define the baseline elastic response
of the BCC lattice. Ordered Ru--X B2 prototype structures were constructed in the CsCl (B2) structure type, with
Ru occupying the A sublattice and a partner element occupying the B sublattice. The primary B2
partners considered in this work were Hf, Ti, Zr, and Al. These binary B2 cells were initialized
with a cubic lattice parameter of approximately 3.1~\AA\ and fully relaxed. In addition to
binary prototypes, a hierarchy of multicomponent B2 and disordered BCC structures was constructed to sample chemically relevant alloying pathways. Figure~\ref{fig:dft_structures} illustrates representative relaxed BCC and B2 structural models
used to construct the DFT dataset. These structures span ordered Ru--X B2 binaries, ternary B2
solubility models, Ru-substituted B2 variants, and disordered BCC parent matrices. Together
they define a consistent structural reference for evaluating formation energetics and lattice
misfit across the Ru-containing refractory alloy space.

Chemically disordered BCC matrices were modeled using special quasirandom structures (SQS)
to approximate random solid solutions. Binary BCC matrices were generated for all pairwise
combinations of Nb, Mo, Ta, W, and V at target compositions of 12.5, 25, 50, 75, and
87.5~at.\% of the first element using $2\times2\times2$ BCC supercells containing 16 atoms.
Equiatomic ternary BCC matrices were generated for Nb--Mo--Ta, Nb--Mo--W, Nb--Mo--V,
Nb--Ta--W, Nb--Ta--V, Nb--W--V, Mo--Ta--W, Mo--Ta--V, Mo--W--V, and Ta--W--V
using $3\times3\times1$ supercells containing 18 atoms, which allow an exact one-third
occupation for each species. For B2-based multicomponent systems, the A and B sublattices were treated explicitly. Ru
always occupied the A sublattice in this work. The B sublattice contained either a single B2
partner or a mixture of a primary B2 former (Hf, Ti, or Zr) and one refractory matrix element
(Nb, Mo, Ta, W, or V). \revB{Ternary B2 solubility structures were constructed by keeping the A sublattice fully occupied by Ru and partially replacing the primary B2 former on the B sublattice with the matrix element, at primary-former fractions of 0.875 and 0.75 on the B sublattice.}
These systems were modeled using $2\times2\times2$ B2 supercells containing eight A sites
and eight B sites. Ru substitution penalty structures were constructed by partially replacing Ru on the A
sublattice with Al, Cr, Cu, or Si at Ru fractions of 1.0, 0.875, 0.75, 0.625, and 0.5, while
keeping the B sublattice fully occupied by Hf, Ti, or Zr. Higher-order B2 structures were also
generated by mixing two B2 partner elements (e.g., Hf--Al or Ti--Zr) in a 50:50 ratio on the B
sublattice, combined with partial Ru--Al mixing on the A sublattice. All supercells and special quasirandom structures were generated with Python using the
\texttt{ase} library for structure construction and the \texttt{icet} framework for SQS generation
\cite{aangqvist2019icet}. The icet ClusterSpace formalism was used to enforce target site occupancies
on either a single BCC sublattice or on the symmetry-distinct A and B sublattices of the B2
(CsCl-type) structure. Monte Carlo optimization was employed to minimize short-range order
subject to exact composition constraints. In addition to serving as thermodynamic reference states, the disordered BCC matrices were
treated as the elastic parent phase for subsequent B2--BCC lattice misfit calculations. The
relaxed lattice parameters of these BCC SQS cells therefore define the matrix lattice constant
$a_{\mathrm{BCC}}$ used in the coherent misfit descriptor. This choice reflects the experimental
observation that Ru-based B2 precipitates form coherently within chemically disordered BCC
matrices rather than within ordered reference compounds. \rev{All structures were fully relaxed with respect to both atomic positions and cell shape using the production settings documented in Supplementary Table~S1 and first-order Methfessel--Paxton smearing with 0.20~eV for the metallic screening calculations. The 16-atom screening cells used a $6\times6\times6$ Monkhorst--Pack mesh; other cell sizes used a comparable reciprocal-space sampling density. All calculations were performed in the non-magnetic state. Static single-point calculations were then carried out on the relaxed geometries with the same electronic convergence target and, where appropriate, tetrahedron integration.}


\subsubsection{\rev{Reference-state convention and temperature scope}}

\rev{Two reference-state conventions are used for distinct purposes in this work. 
The primary BCC--B2 screening dataset uses elemental BCC reference states because 
the design problem concerns ordered B2 precipitates embedded in a BCC parent 
solid solution. This convention provides a common energetic basis for comparing 
ordered B2 structures with BCC-based matrix states throughout the screening 
workflow. Absolute formation enthalpies used for comparison with calorimetric data are 
also evaluated on the elemental ground-state scale. For Ru, Hf, Ti, and Zr, 
matched HCP and BCC calculations were therefore performed using identical 
electronic-structure settings. The corresponding BCC energies lie above the HCP 
energies by 0.620, 0.180, 0.101, and 0.078~eV/atom for Ru, Hf, Ti, and Zr, 
respectively. These energy differences provide the reference correction required 
to convert the BCC-referenced formation enthalpies of RuHf, RuTi, and RuZr to 
the elemental ground-state scale used in the calorimetric comparison of 
Section~3.6. For a composition-matched B2/BCC comparison, the choice of elemental 
reference does not affect the energy difference between the two states. Changing 
the elemental reference shifts both the B2 and BCC formation energies by the 
same composition-weighted constant, which cancels in 
$\Delta E_{\mathrm{B2-BCC}}$. The reference-state choice therefore affects the 
absolute formation enthalpy but not the ordering-energy contrast between 
composition-matched B2 and BCC states. The complete screening space also contains Cartesian pairings between B2 
precipitates and BCC matrices that do not have the same overall composition. 
For these composition-mismatched pairs, $\Delta E_{\mathrm{B2-BCC}}$ is retained 
as an energy-contrast descriptor for candidate screening and ranking rather than 
being interpreted as a thermodynamic reaction energy or precipitation free 
energy. All DFT calculations reported here are static 0~K calculations. Vibrational 
free energies, thermal expansion, finite-temperature electronic free energies, 
and explicit finite-temperature sampling are not included. Temperature enters 
the subsequent symbolic model through the configurational-entropy term and 
through calibration to published solvus data. The resulting 
$T_{\mathrm{solvus}}$ therefore represents an effective calibrated design 
temperature rather than a directly calculated finite-temperature phase 
boundary.}

{\color{red}
\begin{table}[H]
\centering
\caption{Matched elemental HCP and BCC reference-state calculations used to 
convert BCC-referenced formation enthalpies to the elemental ground-state scale. 
The positive values of $\Delta E_{\mathrm{HCP}\rightarrow\mathrm{BCC}}$ show 
that the HCP structure lies below the BCC structure for Ru, Hf, Zr, and Ti 
within the present calculations.}
\label{tab:reference_state_revision}
\begin{tabular}{l l c c c}
\toprule
Element &
Ground-state structure &
$E_{\mathrm{HCP}}$ (eV/atom) &
$E_{\mathrm{BCC}}$ (eV/atom) &
$\Delta E_{\mathrm{HCP}\rightarrow\mathrm{BCC}}$ (eV/atom) \\
\midrule
Ru & HCP & $-9.2039$ & $-8.5843$ & $+0.6196$ \\
Hf & HCP & $-9.9558$ & $-9.7760$ & $+0.1798$ \\
Zr & HCP & $-8.4776$ & $-8.3994$ & $+0.0782$ \\
Ti & HCP & $-7.7628$ & $-7.6615$ & $+0.1013$ \\
\bottomrule
\end{tabular}
\end{table}
}

\subsection{Descriptor construction and rule-of-mixtures features}

Each alloy configuration was represented using a set of physically
motivated descriptors derived directly from density functional theory.
The descriptor set was designed to capture both the thermodynamic
driving force for B2 formation and the elastic compatibility between
ordered B2 precipitates and disordered BCC matrices. All descriptors
were constructed from relaxed DFT quantities or from compositionally
resolved data derived from the DFT dataset. No empirical or proxy misfit
parameters were used.

Thermodynamic stability was characterized using DFT formation energies.
For each B2 structure, a per-atom formation energy
$E_{\mathrm{form}}^{\mathrm{B2}}$ was computed relative to the relaxed
elemental BCC reference states. An analogous formation energy
$E_{\mathrm{form}}^{\mathrm{BCC}}$ was computed for each disordered BCC
matrix supercell. A driving-force descriptor for B2 precipitation was
then defined as the formation-energy difference
\begin{equation}
\Delta E_{\mathrm{B2-BCC}} =
E_{\mathrm{form}}^{\mathrm{B2}} - E_{\mathrm{form}}^{\mathrm{BCC}},
\label{eq:driving_force}
\end{equation}
\rev{For composition-matched B2 and BCC states, this quantity is an ordering-energy difference at fixed composition. In the broader Cartesian matrix--precipitate screen, many paired rows do not have identical overall composition; in those cases the same numerical difference is retained only as an energy-contrast descriptor for screening and is not interpreted as a thermodynamic precipitation driving force.} Elastic coherency between B2 precipitates and BCC matrices was
quantified using a lattice-parameter-based misfit descriptor derived
directly from relaxed DFT lattice constants. For each Ru-based B2
structure, the relaxed cubic lattice parameter $a_{\mathrm{B2}}$ was
extracted. For each BCC matrix composition, the relaxed lattice
parameter $a_{\mathrm{BCC}}$ was obtained from the corresponding BCC
SQS calculation. The coherent lattice misfit was defined using the
symmetric form
\begin{equation}
\delta =
\frac{2\left(a_{\mathrm{B2}} - a_{\mathrm{BCC}}\right)}
     {a_{\mathrm{B2}} + a_{\mathrm{BCC}}},
\label{eq:lattice_misfit}
\end{equation}
which is widely used in experimental and computational studies of
coherent BCC--B2 systems. Equation~\eqref{eq:lattice_misfit} treats the
BCC matrix as the elastic parent phase and captures both the magnitude
and sign of the misfit, which are relevant for elastic strain energy,
coherency loss, and precipitate morphology.

To systematically evaluate matrix--precipitate compatibility, each B2
structure was paired with every BCC matrix composition, forming a
cartesian product of B2--BCC combinations. The misfit $\delta$ defined
in Eq.~\eqref{eq:lattice_misfit} was evaluated for every pair using their
respective relaxed lattice parameters. This approach enables direct
comparison of a given B2 phase across a wide range of candidate BCC
matrices and mirrors the design logic employed in experimental studies
of Ru-based B2 precipitates. Complementing the DFT-derived energetics and misfit, several composition-dependent descriptors were calculated using the linear rule of mixtures to capture electronic and chemical trends. These included the valence electron concentration ($\mathrm{VEC}_{\mathrm{B2}}$), the average Pauling electronegativity ($\bar{\chi}_{\mathrm{B2}}$), and the electronegativity difference ($\Delta \chi_{\mathrm{B2}}$) across the B2 sublattices. Structural disorder was parameterized using the atomic size mismatch $\delta_r = \sqrt{\sum c_i (1 - r_i/\bar{r})^2}$, while chemical interactions were approximated using the mixing enthalpy ($H_{\mathrm{mix}}$) based on Miedema's model \cite{zhang2016miedema}. We deliberately restricted the rule-of-mixtures approximation to this compact set of fundamental physical parameters, avoiding the proliferation of excessive heuristic features. All descriptors were standardized prior to model training using z-score normalization based on the mean and standard deviation of the training dataset.


\subsection{Data filtering and ML model training}

\begin{figure}[H]
    \centering
    \includegraphics[width=\textwidth]{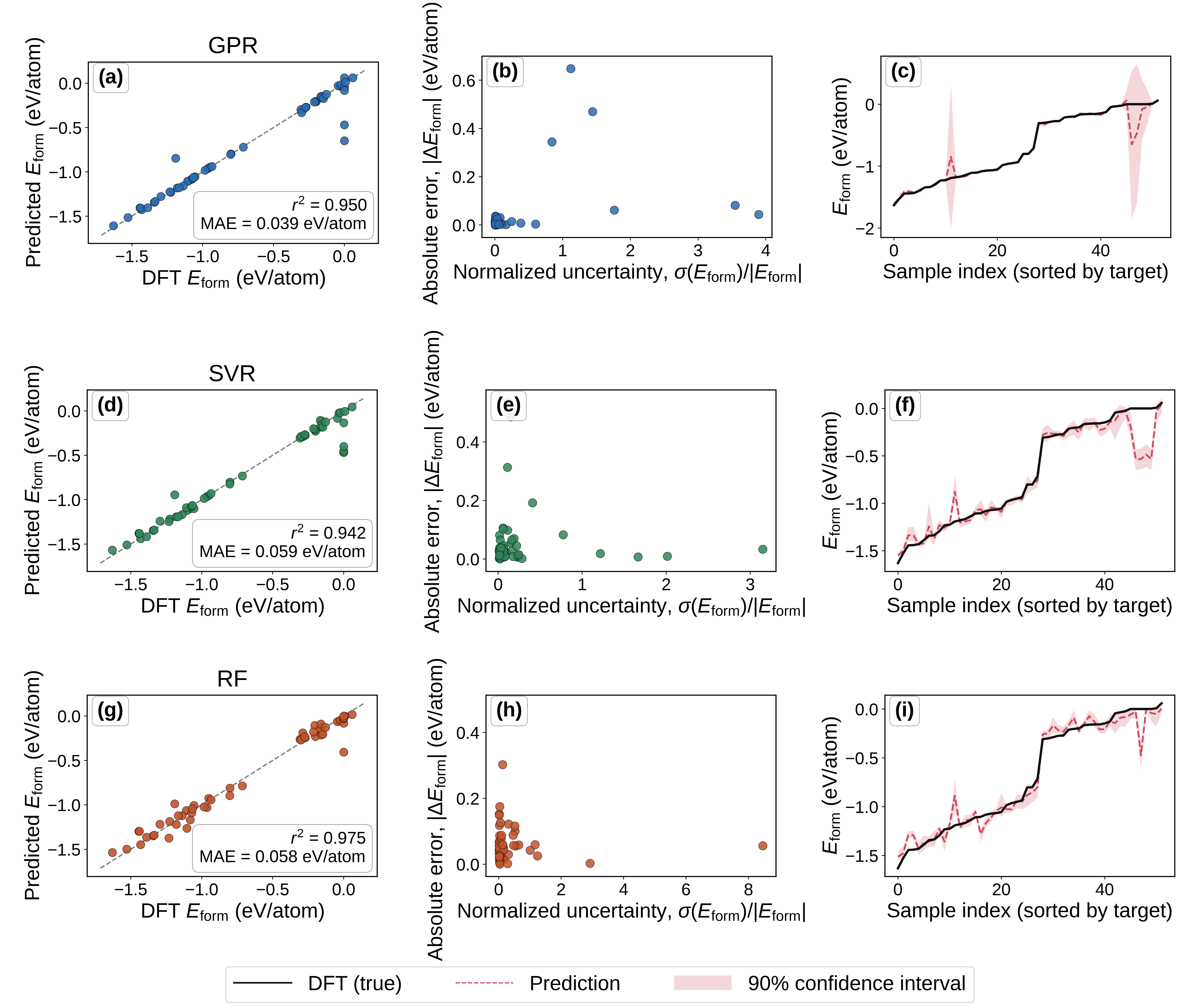}
    \caption{\rev{Comparison of Gaussian process regression (GPR), support vector regression (SVR), and random forest (RF) models for predicting DFT formation energies. Parity between DFT formation energies and model predictions is shown for (a) GPR, (d) SVR, and (g) RF. The relationship between absolute prediction error and normalized predictive uncertainty, $\sigma(E_{\mathrm{form}})/|E_{\mathrm{form}}|$, is shown for (b) GPR, (e) SVR, and (h) RF. DFT values together with the corresponding model predictions are shown for (c) GPR, (f) SVR, and (i) RF after sorting the samples by target formation energy; the shaded regions represent the 90\% confidence intervals. Formation energies and absolute errors are reported in eV/atom. Predictive uncertainty is represented directly on the horizontal axis.}}
    \label{fig:ml_figures}
\end{figure}

Three regression models were evaluated to learn the mapping between DFT-derived descriptors and formation-energy-based stability metrics: Gaussian process regression (GPR) \cite{deringer2021gaussian}, support vector regression (SVR) \cite{awad2015support}, and random forest regression (RFR) \cite{rodriguez2015machine1}. \rev{All three models were evaluated using the same filtered dataset and the same feature set, allowing their predictive performance and uncertainty behavior to be compared directly.} Fig.~\ref{fig:ml_figures} summarizes the comparative performance of the three models. Fig.~\ref{fig:ml_figures}(a), (d), and (g) show parity plots between DFT formation energies and model predictions for Gaussian process regression, support vector regression, and random forest regression, respectively. All three models achieve high predictive accuracy. Gaussian process regression yields a mean absolute error of approximately 0.04~eV/atom with an $r^2$ value of about 0.95 (Fig.~\ref{fig:ml_figures}(a)). Support vector regression shows slightly larger scatter, with a mean absolute error near 0.06~eV/atom and $r^2\approx0.94$ (Fig.~\ref{fig:ml_figures}(d)). Random forest regression achieves the highest overall $r^2$ value of approximately 0.97, with a mean absolute error comparable to SVR at about 0.06~eV/atom (Fig.~\ref{fig:ml_figures}(g)).

Uncertainty behavior was analyzed to assess model robustness. Fig.~\ref{fig:ml_figures}(b), (e), and (h) show scatter plots of absolute prediction error versus normalized predictive uncertainty for GPR, SVR, and RFR, respectively. Gaussian process regression exhibits a clear correlation between predicted uncertainty and actual error (Fig.~\ref{fig:ml_figures}(b)), indicating that larger model uncertainty generally coincides with larger prediction errors. Support vector regression and random forest regression use resampling-based uncertainty estimates. In both cases, the normalized uncertainty remains small for most samples, while isolated points with larger errors occur in more sparsely represented regions of the dataset (Fig.~\ref{fig:ml_figures}(e), (h)). Fig.~\ref{fig:ml_figures}(c), (f), and (i) show uncertainty-aware predictions plotted against the sorted target values. For all three models, the predicted mean follows the DFT trend closely across the full formation-energy range. Gaussian process regression produces smooth confidence intervals but shows occasional large uncertainty excursions near sparsely sampled portions of the data distribution (Fig.~\ref{fig:ml_figures}(c)). Support vector regression produces broader confidence intervals in several regions (Fig.~\ref{fig:ml_figures}(f)). Random forest regression yields comparatively narrow confidence intervals across most of the dataset, with localized widening where the underlying data are sparse (Fig.~\ref{fig:ml_figures}(i)).

\rev{Among the three models, Gaussian process regression gives the lowest mean absolute error, whereas random forest regression provides the highest overall $r^2$. The random-forest model is retained for the subsequent design-space analysis because its ensemble architecture provides a practical surrogate for nonlinear interactions among the thermodynamic, compositional, and structural descriptors and can be applied efficiently across the full candidate library. The comparison with GPR and SVR therefore serves both as a model-performance benchmark and as a consistency check that the principal formation-energy trends are not specific to a single regression architecture.}

\subsection{Physics-guided symbolic regression of solvus temperature with misfit constraints}

The stability of ordered B2 precipitates in refractory BCC alloys arises from a balance between thermodynamic driving forces for ordering and elastic penalties associated with lattice misfit. In this work, we formulate the solvus temperature as a physics-guided regression problem. The dominant contribution comes from a DFT-derived enthalpic scale and the configurational entropy of mixing. Lattice misfit provides an additional constraint that limits the stability of coherent B2 phases.

To ensure that the model reflects these physical considerations, we construct descriptors directly from first-principles calculations. These include the absolute formation-energy difference between BCC and B2 phases, the exact ideal configurational entropy computed from alloy compositions, and the BCC–B2 lattice misfit. This approach captures the main features of mean-field order–disorder theory while allowing elastic effects to modify the predictions. Symbolic regression (SR) is then applied in a staged and physics-guided manner to derive an interpretable solvus relation.

\rev{Two energetic quantities used in the workflow must be distinguished. The pairwise quantity $\Delta E_{\mathrm{B2-BCC}}$ defined in Eq.~\eqref{eq:driving_force} is used as a screening descriptor when B2 structures are paired with candidate BCC matrices. By contrast, $\Delta H$ in the solvus relation below denotes the positive magnitude of the formation enthalpy of the B2 precipitate, $|E_{\mathrm{form}}^{\mathrm{B2}}|$. The entropy term is the ideal configurational entropy of the corresponding disordered BCC matrix, $\Delta S_{\mathrm{mix}}=-k_{\mathrm B}\sum_i c_i\ln c_i$. Thus, for a fixed B2 precipitate, $\Delta H$ is fixed, whereas $\Delta S_{\mathrm{mix}}$ changes only with matrix composition. This distinction is important because $\Delta E_{\mathrm{B2-BCC}}$ and $\Delta H$ serve different roles in the screening and solvus-calibration stages of the workflow.}

We first examine a purely thermodynamic expression,
\begin{equation}
T^{(0)}_{\mathrm{solvus}} = a\,\frac{\Delta H}{\Delta S_{\mathrm{mix}}},
\label{eq:solvus0}
\end{equation}
which represents the leading-order enthalpy–entropy balance. This expression consistently overestimates stability for alloys that possess large lattice mismatch, and the limitation motivates the introduction of a misfit penalty. We therefore modify the target to
\begin{equation}
T^{(1)}_{\mathrm{solvus}} = a\,\frac{\Delta H}{\Delta S_{\mathrm{mix}}} - b\,\delta,
\label{eq:solvus1}
\end{equation}
\rev{where $\delta$ is the absolute BCC--B2 misfit and the constants $a$ and $b$ calibrate the relation to the literature solvus dataset. Because $\Delta H/\Delta S_{\mathrm{mix}}$ has units of kelvin, the symbolic-regression expression is evaluated first as an effective temperature in kelvin; values reported in degrees Celsius are obtained using $T(^\circ\mathrm{C})=T(\mathrm{K})-273.15$. The use of temperature in this calibrated relation does not imply that the underlying DFT calculations were performed at finite temperature.}

To prevent the symbolic regression from producing unphysical expressions, we restrict the operator set to simple algebraic forms and limit the feature space to thermodynamic and misfit-related quantities. We then provide higher-order misfit terms and strain-energy-like combinations,
\begin{equation}
T_{\mathrm{solvus}} = 
f\!\left(
\frac{\Delta H}{\Delta S_{\mathrm{mix}}},\ 
\delta,\ 
\delta^2,\ 
\frac{\delta^2}{\Delta S_{\mathrm{mix}}}
\right),
\label{eq:solvus_f}
\end{equation}
so that the algorithm can refine the relation in a controlled manner. The resulting expressions show a consistent pattern: the thermodynamic ratio forms the backbone of the solvus temperature, and the misfit terms act as subtractive corrections whose influence increases with lattice strain. This progression shows that thermodynamics alone cannot explain high-temperature B2 stability, and that elastic compatibility must also be included.

\noindent Symbolic expressions serve as interpretable filters for alloy design. We explore the design space by screening compositions using both the predicted solvus temperature and the lattice misfit. In this way we identify Ru-containing B2 systems that combine strong thermodynamic stabilization with acceptable coherency against the parent BCC matrix. Solvus temperature and misfit therefore operate as coupled design constraints rather than independent metrics. This combined strategy is consistent with recent symbolic-regression studies that seek compact analytical relations for other temperature-dependent material transitions, and it provides a clear route for narrowing the refractory-alloy search space to compositions that satisfy energetic, entropic, and elastic requirements for B2 stability.


\section{Results}

\subsection{DFT formation energetics of BCC and B2 structures}
\label{subsec:dft_formation}

Density functional theory calculations establish a consistent energetic baseline for comparing the stability of chemically disordered BCC matrices and ordered Ru based B2 structures. All formation energies are reported on a per atom basis relative to the elemental BCC reference states defined in Section~2.1. \rev{The BCC cells of Nb, Mo, Ta, W, V, Hf, Zr, and Ti all converged within the imposed BCC structure and provide a consistent internal reference set; this should not be read as a statement that Hf, Zr, or Ti are ambient-pressure BCC ground states. The targeted HCP--BCC reference checks introduced above make that distinction explicit.} Their equilibrium lattice parameters and cohesive trends agree with established first principles benchmarks. Binary and multicomponent BCC matrices show smooth and monotonic variation of formation energy with composition. This behavior indicates that the structure generation and relaxation workflow does not introduce artificial ordering. Increasing the concentration of Mo or W lowers the BCC formation energy because these elements have high cohesive energies. Increasing the concentration of V or Ta raises the formation energy slightly while preserving overall stability. Representative examples from the BCC and B2 dataset are listed in Table~\ref{tab:dft_table_representative}. \revB{The complete dataset is available in the project repository listed in the Data availability statement, and the extended list of ideal candidates is given in Supplementary Table~S16.}

{\color{red}
\begin{table}[H]
\centering
\caption{\rev{Representative DFT formation energies and relaxed lattice parameters for selected BCC matrices and Ru-based B2 structures. Formation energies are referenced to the elemental BCC states used in the screening dataset.} \revB{$\Delta E_{\mathrm{form}}^{\mathrm{BCC\ ref}}$ is the formation energy of the phase listed in the Structure Type column (the ordered B2 structure or the disordered BCC SQS matrix) relative to its constituent pure elements, each taken in the BCC structure, i.e., $\Delta E_{\mathrm{form}}^{\mathrm{BCC\ ref}}=E_{\mathrm{phase}}-\sum_i c_i E_i^{\mathrm{BCC}}$. It is not a B2--BCC energy difference; the pairwise descriptor $\Delta E_{\mathrm{B2-BCC}}$ of Eq.~\ref{eq:driving_force} is evaluated separately for each B2--BCC pair.}}
\label{tab:dft_table_representative}
\begin{tabular}{l l c c}
\hline
System & Structure Type & $\Delta E_{\mathrm{form}}^{\mathrm{BCC\ ref}}$ (eV/atom) & Lattice Parameter (\AA) \\
\hline
HfRu & B2 binary & $-1.190$ & $3.239$ \\
ZrRu & B2 binary & $-0.965$ & $3.269$ \\
TiRu & B2 binary & $-1.053$ & $3.074$ \\
RuAl & B2 binary & $-1.009$ & $3.009$ \\
\hline
(Hf,Mo)Ru & B2 with B-site mixing & $-1.011$ & $3.218$ \\
(Zr,Ta)Ru & B2 with B-site mixing & $-0.921$ & $3.258$ \\
\hline
(Ru$_{0.75}$Al$_{0.25}$)--Hf & B2 with A-site substitution & $-0.926$ & $3.263$ \\
Ru$_{0.875}$Cr$_{0.125}$--Hf & B2 with A-site substitution & $-1.035$ & $3.229$ \\
Ru$_{0.875}$Si$_{0.125}$--Hf & B2 with A-site substitution & $-1.077$ & $3.243$ \\
\hline
Nb$_{50}$Mo$_{50}$ & BCC matrix (binary) & $-0.081$ & $3.231$ \\
Mo$_{50}$Ta$_{50}$ & BCC matrix (binary) & $-0.118$ & $3.222$ \\
\hline
\end{tabular}
\end{table}
}

\begin{figure}[H]
    \centering

    \includegraphics[width=0.75\textwidth]{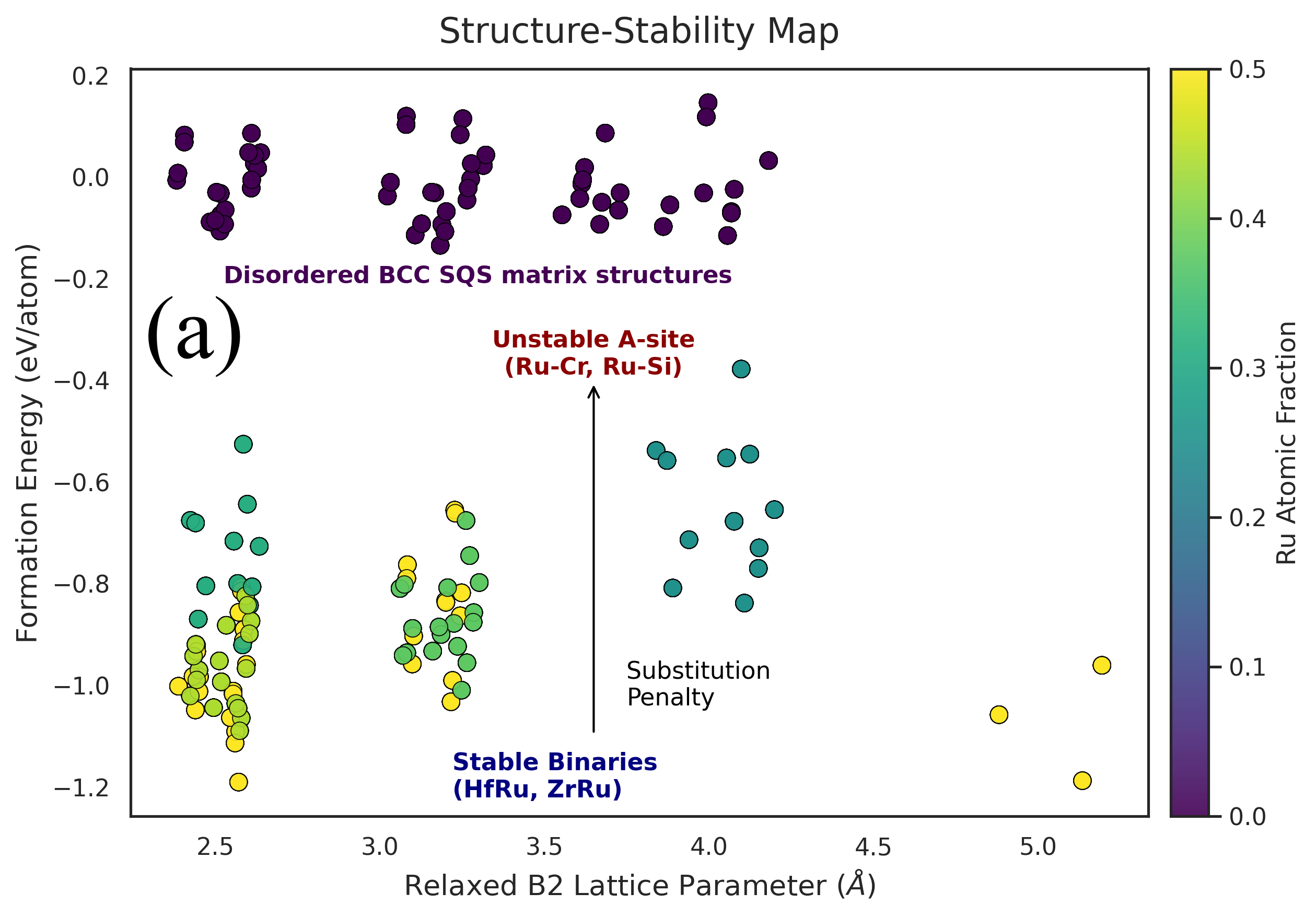}


    \includegraphics[width=0.75\textwidth]{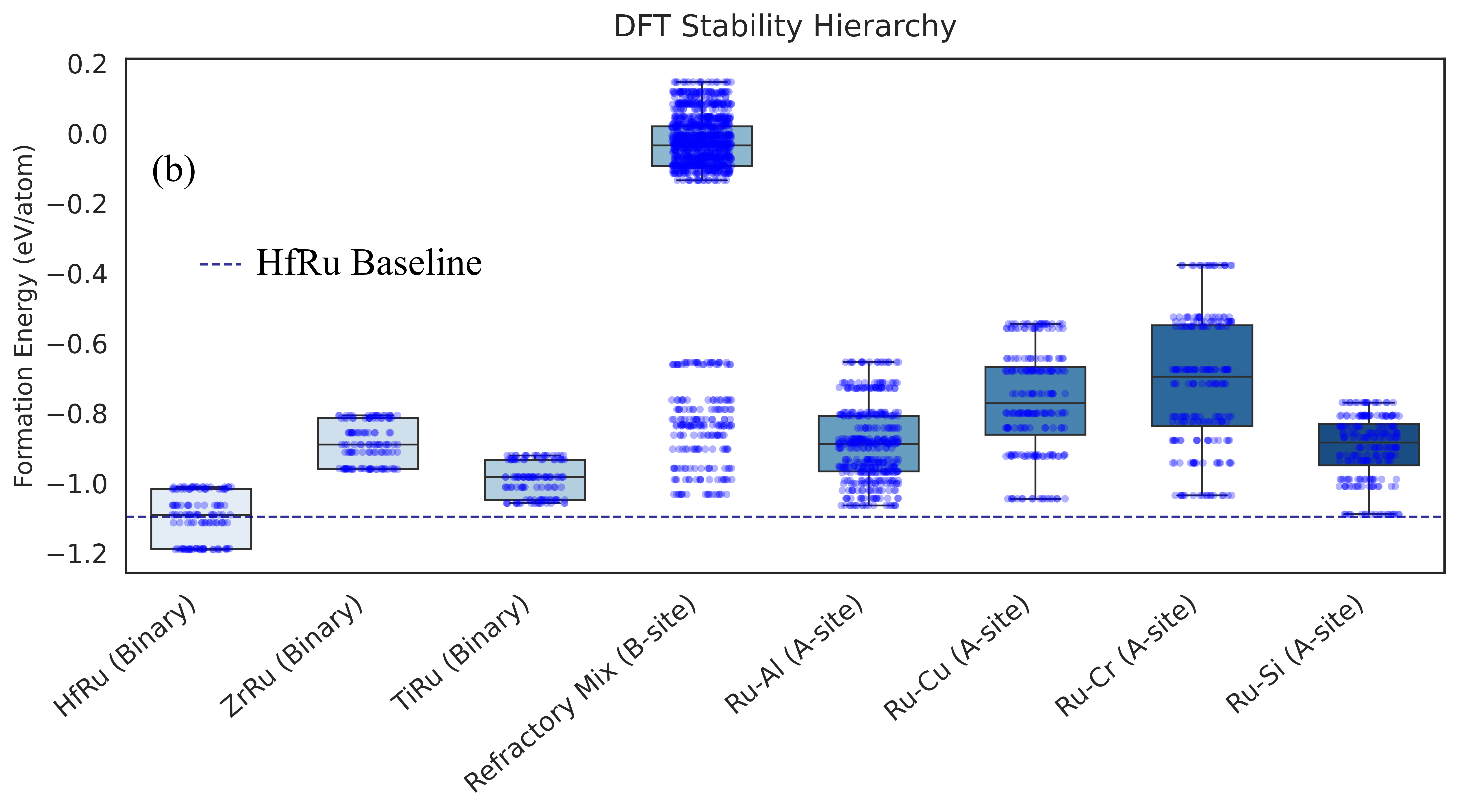}

    \caption{\rev{DFT formation energetics of Ru-based B2 structures. 
    (a) Structure--stability map showing the DFT formation energy as a function of the relaxed B2 lattice parameter, with points colored by Ru atomic fraction. Strongly bound Ru-rich binaries such as HfRu and ZrRu occupy the low-energy region, whereas chemical substitution generally shifts the formation energy toward zero. 
    (b) Formation-energy distributions grouped by chemical class. The Ru-based binaries provide the stability baseline, while refractory B-site mixing and Ru-site substitution progressively reduce the magnitude of the formation energy. The dashed line indicates the HfRu binary reference energy.}}
    \label{fig:dft_energies}
\end{figure}

Table~\ref{tab:dft_table_representative} highlights the major stability trends observed across the full dataset. The Ru based B2 binaries show the most negative formation energies and relaxed lattice parameters in the range of 3.1--3.3~\AA. These systems represent the strongest ordering tendency and therefore provide the natural baseline for evaluating destabilization caused by alloying. \rev{B2 structures containing a mixture of a primary B2 former and a refractory matrix element on the B site remain negative in formation energy but generally move toward zero, i.e., the magnitude $|\Delta E_{\mathrm{form}}|$ decreases. B2 structures with partial substitution of Ru on the A site by Al, Cu, Cr, or Si show a larger reduction in $|\Delta E_{\mathrm{form}}|$.} These representative values capture the same hierarchy observed in the full dataset. To visualize these energetic trends in more detail, Fig.~\ref{fig:dft_energies}(a) shows a structure stability map where the horizontal axis corresponds to the relaxed B2 lattice parameter and the vertical axis corresponds to the DFT formation energy. \rev{The near-zero-energy points at zero Ru fraction correspond to the disordered BCC SQS matrix structures and are not members of the Ru-substituted B2 series. The Ru--Cr and Ru--Si designations refer specifically to Ru-containing A-sublattice-substituted B2 configurations.} Each point is colored by the Ru atomic fraction on the A sublattice. The stable Ru rich binaries cluster near formation energies of approximately $-1.0$~eV/atom and lattice parameters between 3.1 and 3.3~\AA{}. This region defines the chemical and structural window that favors strong B2 ordering. Systems with larger lattice parameters or higher formation energies correspond to B site mixed or A site substituted chemistries. These points form a clear upward trend that reflects the substitution induced destabilization. The annotation labelled ``Substitution Penalty'' in Fig.~\ref{fig:dft_energies}(a) highlights this systematic energetic shift. Figure~\ref{fig:dft_energies}(b) further summarizes the formation energy distributions by chemical category. The categories ``HfRu (Binary)'', ``ZrRu (Binary)'', and ``TiRu (Binary)'' correspond to the ideal Ru based B2 binaries that define the stability baseline. The category ``Refractory Mix (B-site)'' contains structures where the B sublattice includes a mixture of a primary B2 former and a refractory element such as Nb, Mo, Ta, or W. \rev{These systems remain negative in formation energy but show a moderate reduction in $|\Delta E_{\mathrm{form}}|$.} The categories ``Ru-Al (A-site)'', ``Ru-Cu (A-site)'', ``Ru-Cr (A-site)'', and ``Ru-Si (A-site)'' correspond to A site substitution of Ru. These classes show progressively higher formation energies, with Cr and Si substitution producing the largest destabilizing effects. Together these results establish a clear hierarchy of B2 stability that spans BCC matrices, Ru rich B2 binaries, B site mixed B2 structures, and A site substituted B2 variants. This hierarchy provides the thermodynamic foundation for the lattice misfit descriptors and symbolic regression model developed in the sections that follow. \rev{The systematic reduction in $|\Delta E_{\mathrm{form}}|$ with chemical substitution supports the interpretation of a reduced thermodynamic contribution to the effective B2 solvus temperature in Ru-lean compositions.}

\subsection{Interplay between thermodynamic stability and lattice misfit}
\label{subsec:misfit}

\begin{figure}[H]
\centering
\includegraphics[width=0.75\textwidth]{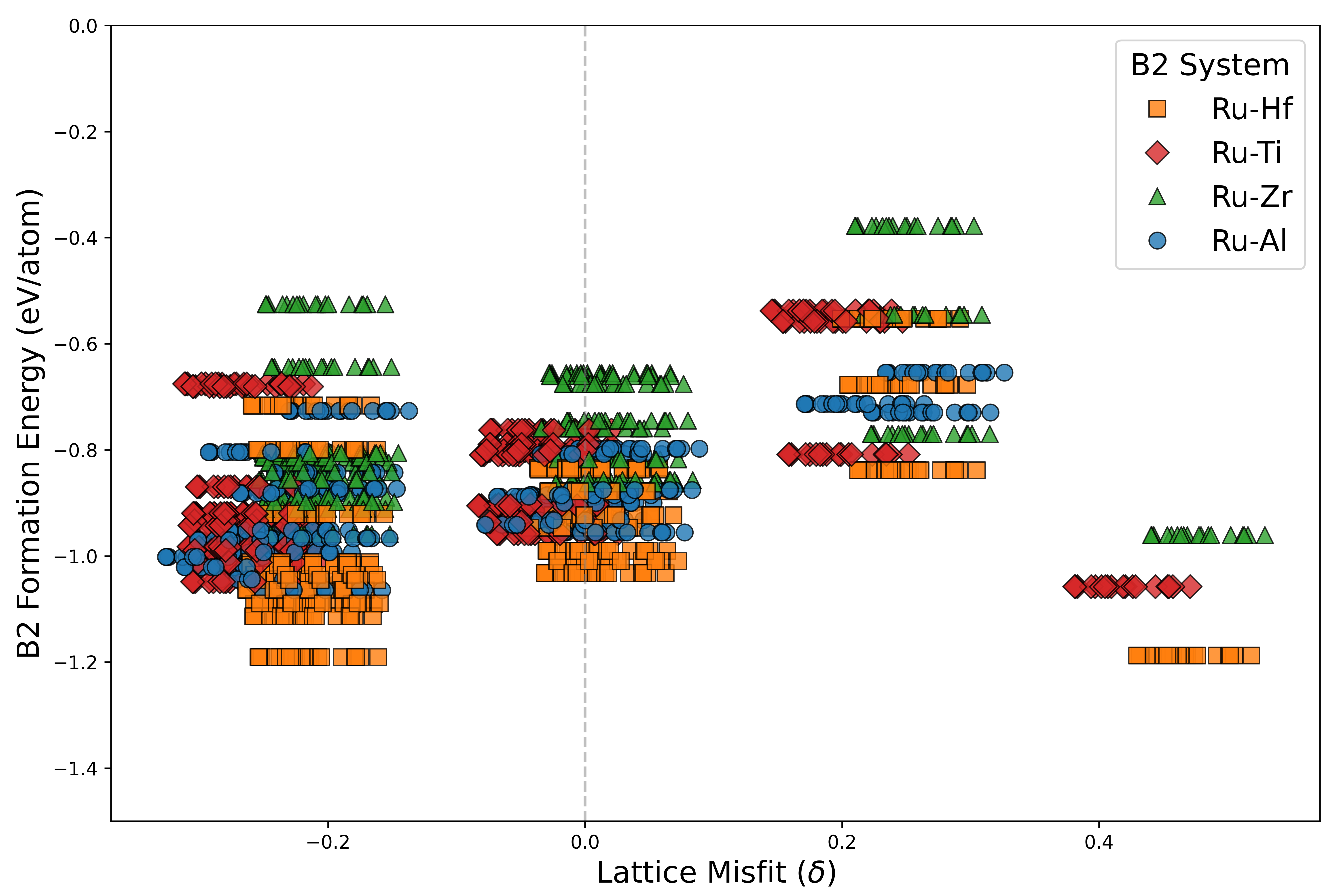}
\caption{\rev{Relationship between B2 formation energy and coherent lattice misfit $\delta$ for representative Ru--Hf, Ru--Ti, Ru--Zr, and Ru--Al B2/BCC combinations. Each B2 chemical family spans multiple candidate BCC matrix pairings and therefore exhibits a distribution of misfit values. Several Ru--Al combinations occur near the zero-misfit condition, demonstrating that coherency is a property of the individual matrix--precipitate pair rather than of the B2 family alone. The dashed vertical line marks $\delta=0$.}}
\label{fig:dft_lattice_misfit}
\end{figure}

\rev{The stability of Ru-based B2 precipitates in refractory matrices depends not only on their intrinsic formation energies but also on their lattice coherency with the surrounding BCC solid solution. To quantify this interplay, Fig.~\ref{fig:dft_lattice_misfit} plots the DFT-calculated B2 formation energies against the corresponding coherent lattice misfit $\delta$ for representative Ru--Hf, Ru--Zr, Ru--Ti, and Ru--Al systems.} \rev{The coherent misfit is determined by the specific B2 precipitate--BCC matrix pairing rather than by the B2 chemical family alone. Each B2 chemistry therefore spans a range of $\delta$ values as it is paired with different candidate BCC matrices. In particular, the Ru--Al systems include several combinations close to the zero-misfit condition and overlap the low-misfit region sampled by the Ru--Hf and Ru--Zr families. Ru--Al should therefore not be interpreted as a uniformly incoherent B2 family. The distribution of $|\delta|$ for each chemical family, including the median, interquartile range, and fractions falling within the $\pm1\%$ and $\pm2\%$ coherency windows, is summarized in Supplementary Table~S9.} \rev{The Ru--Al chemical family shown in Fig.~\ref{fig:dft_lattice_misfit} is distinct from the use of Al as a substitutional lattice-tuning addition in Ru--Hf-based B2 phases. In the former case, Ru--Al defines the B2 chemistry itself, whereas in the latter case partial replacement of Ru by Al is used to modify the lattice parameter of an existing Ru--Hf B2 phase. The concentration dependence of this lattice-tuning mechanism and its associated thermodynamic penalty are examined separately in Section~3.5.2.} \rev{Figure~\ref{fig:dft_lattice_misfit} shows the unsmoothed DFT energy--misfit pairs before surrogate modelling. It therefore provides a direct view of the simultaneous energetic and geometric constraints acting on each B2--BCC combination. Stable precipitate candidates require both a sufficiently negative formation energy and a sufficiently small lattice mismatch, but neither quantity alone defines the preferred candidate. The machine-learning analysis in the following section uses these DFT-derived energetic and structural descriptors to explore the broader design space quantitatively.}

\subsection{Machine-learning optimization of the BCC--B2 design space}
\label{sec:ML_optimization}

\begin{figure}[H]
    \centering
    \includegraphics[width=1\textwidth]{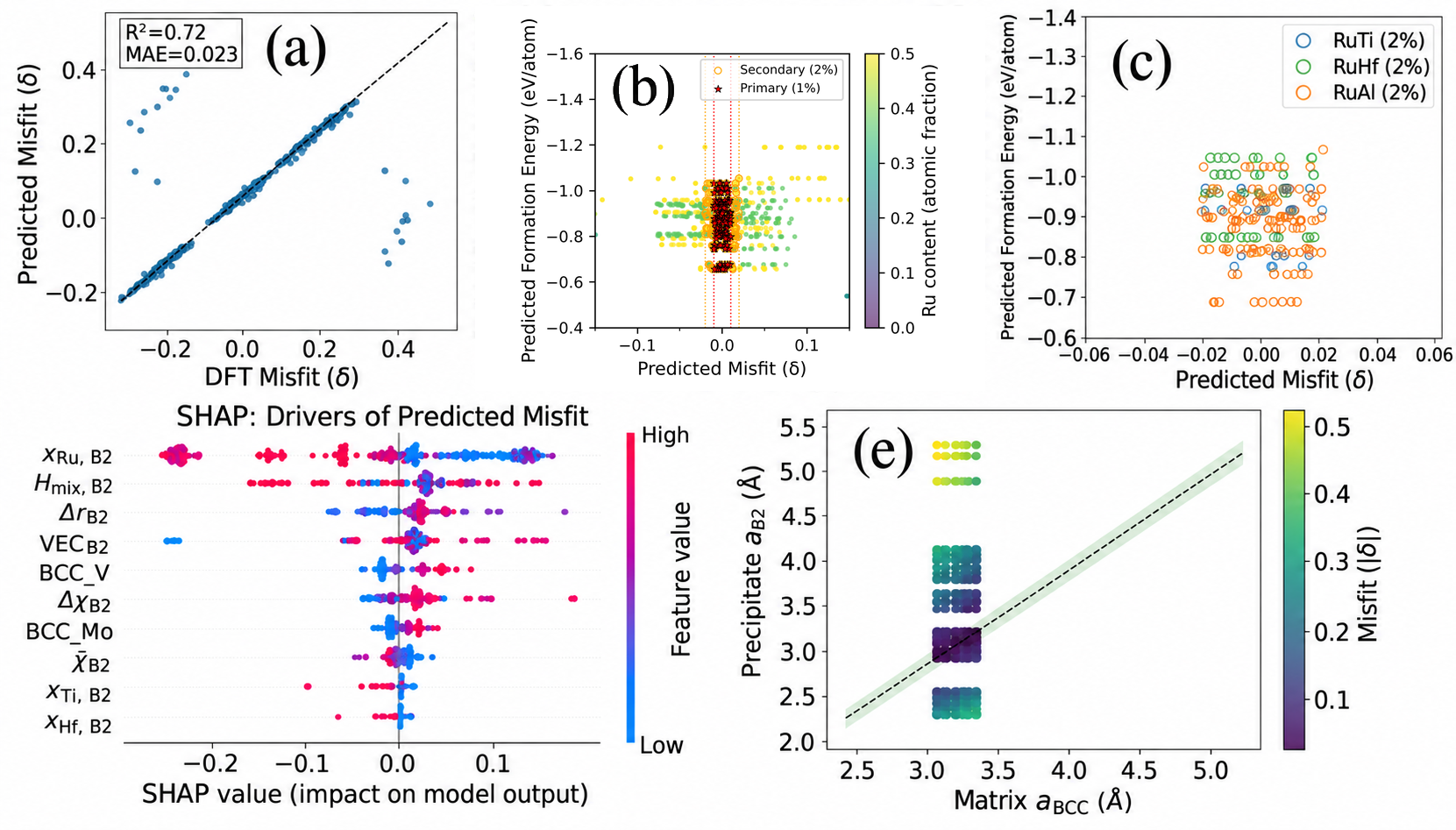}
    \caption{
    Machine-learning optimization of the BCC–B2 design space.  
    (a) RF parity plot for DFT misfit.  
    (b) RF-predicted formation energy vs.\ misfit with $\pm 1\%$ and
    $\pm 2\%$ coherency windows.  
    (c) Branch-wise comparison of RuTi, RuHf, and RuAl systems.  
    (d) SHAP feature-importance ranking for misfit.  
    (e) DFT lattice-matching map showing $a_{\mathrm{B2}}$ vs.\
    $a_{\mathrm{BCC}}$ with coherence band.}
    \label{fig:ML_design}
\end{figure}

\rev{Machine learning was used to extend the DFT-based screening to a larger set of candidate B2--BCC combinations. The random-forest (RF) model described in Section~2.3 was selected for this analysis because it reproduces the DFT trends with high accuracy while allowing rapid evaluation of the full candidate space. Figure~\ref{fig:ML_design}(a) shows the parity between the DFT-calculated and RF-predicted lattice misfit. The predictions follow the DFT values closely over the sampled range, showing that the model captures the principal variation in geometric compatibility among the candidate matrix--precipitate pairs. Figure~\ref{fig:ML_design}(b) shows the corresponding formation-energy--misfit design map. The highest-ranked candidates occur near the low-misfit region while maintaining favorable formation energies. The $\pm1\%$ and $\pm2\%$ bands are used as geometric compatibility windows for identifying coherent and near-coherent candidates. These limits are screening criteria rather than phase-boundary conditions; a small calculated misfit alone does not establish precipitate formation or coherency retention at finite temperature. Figure~\ref{fig:ML_design}(c) shows how the Ru--Ti, Ru--Hf, and Ru--Al chemical families populate this design space. Each family spans a range of misfit values because the misfit depends on the specific B2--BCC pairing rather than on the B2 chemistry alone. The corresponding family-level misfit statistics are given in Supplementary Table~S9. The SHAP analysis in Fig.~\ref{fig:ML_design}(d) identifies the descriptors that contribute most strongly to the RF predictions. These values are used to interpret the response of the trained model and are not assigned directly to a microscopic bonding mechanism. Figure~\ref{fig:ML_design}(e) provides a direct comparison of the relaxed B2 and BCC lattice parameters. Candidate pairs close to the diagonal have similar lattice parameters and therefore small geometric misfit, whereas increasing separation from the diagonal corresponds to progressively larger mismatch. The RF screening therefore narrows the candidate space to combinations that satisfy both energetic and geometric requirements. These candidates are subsequently evaluated using the symbolic-regression relation developed in the following section.}

\begin{figure}[H]
    \centering
    \includegraphics[width=0.9\textwidth]{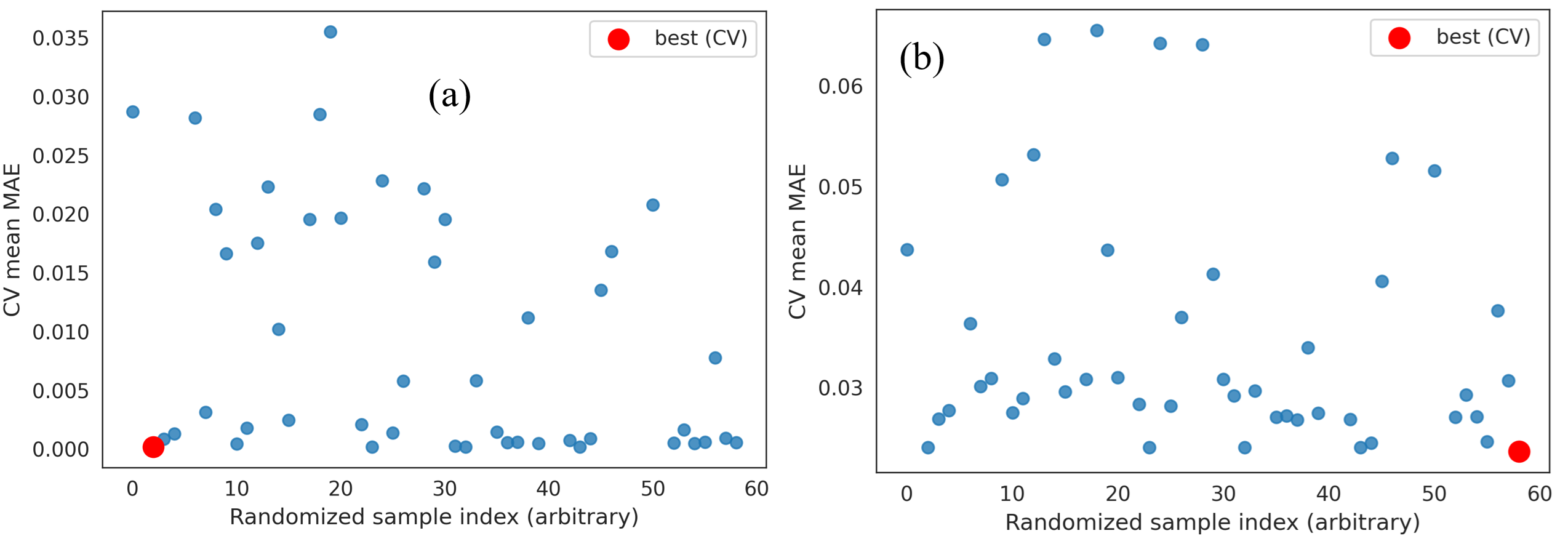}
    \caption{Grid-search validation of RF hyperparameter choices. (a) Cross-validated mean absolute error for formation energy over randomized sample draws. (b) Cross-validated mean absolute error for misfit. The red marker denotes the best cross-validation sample selected for the tuned RF used in this work.}
    \label{fig:ML_grid}
\end{figure}

\noindent The RF settings used to produce Fig.~\ref{fig:ML_design}(a,b) were selected to balance predictive accuracy and stability across energy and misfit targets, and their performance under randomized cross validation is summarized in Fig.~\ref{fig:ML_grid}. As shown in Fig.~\ref{fig:ML_design}(a) the tuned RF reproduces DFT misfit with high fidelity, and Fig.~\ref{fig:ML_design}(b) demonstrates that the same model locates a compact, physically meaningful design window near zero misfit. In Fig.~\ref{fig:ML_grid}(a,b) the scattered distribution of cross-validated errors reflects the variability that arises when the RF ensemble size, tree depth and feature sampling are perturbed. The tuned model is located at the lower bound of these distributions, and most sampled configurations fall within similar error ranges, indicating that the overall structure of the misfit and formation-energy landscape is robust to moderate hyperparameter changes. The small fluctuations seen across the grid search motivate the use of both the primary ($\pm 1\%$) and secondary ($\pm 2\%$) design windows in Fig.~\ref{fig:ML_design}(b), which ensures that conclusions regarding coherent and near-coherent candidate regions remain stable under reasonable model perturbations. We therefore report the grid-search diagnostics in the Supplementary Information and include the tuned model artifacts and CSV outputs as supplementary material; nevertheless, it should be noted that small shifts in model settings or adopting a different algorithm family (for example XGBoost, Bayesian optimization or SVR) and, more importantly, changing the matrix chemistries under consideration, can shift the numerical boundaries of the design windows. For these reasons we rely on the physics-derived descriptors and consistency checks against relaxed DFT lattice parameters when interpreting the RF-guided map; the next section uses symbolic regression to extract compact analytic relations that summarize those physics-constrained trends and to test their stability across resampling.

\subsection{Symbolic Regression Formulation of Coherent Stability}

The RF surrogate developed in Section~3.3 identifies a narrow and physically
meaningful region of the BCC--B2 design space in which strong thermodynamic driving force
coexists with low lattice misfit. In particular, the RF model reveals that high solvus stability is
confined to a compact coherency-controlled domain, consistent with experimental observations
of coherent Ru-based B2 precipitation. While the RF reliably maps this stability landscape, its
ensemble-based architecture does not yield an explicit functional relationship between
thermodynamic stabilization and elastic compatibility. To extract such a relationship and to
rationalize the structure of the RF-predicted stability surface, we apply SR
to the RF-screened dataset. The SR analysis follows the physics-guided formulation introduced in Section~2.4. The regression
was constrained to operate on descriptors with direct thermodynamic or elastic meaning, namely
the enthalpy--entropy ratio $\Delta H/\Delta S_{\mathrm{mix}}$ and the coherent lattice misfit $\delta$.
To ensure that higher-order elastic effects were not artificially excluded, the feature set explicitly
included nonlinear misfit descriptors such as $\delta^2$, $\delta^3$, and strain-energy-like
combinations (e.g., $\delta^2/\Delta S_{\mathrm{mix}}$). Unary operators allowing squaring and
cubing were enabled during the search, and parsimony penalties were deliberately kept low,
allowing complex expressions to persist where they provided meaningful improvement in
predictive fidelity.

Across extensive symbolic regression runs and inspection of the hall-of-fame solutions, the
regression consistently converged to a form that explicitly decouples thermodynamic and elastic
contributions. \rev{The resulting expression, given in Eq.~(6), expresses the effective solvus design temperature as a direct competition between a thermodynamic ordering term and a coherency penalty:}
\begin{equation}
\rev{T_{\mathrm{SR}}(\mathrm{K}) =
0.1121\left(\frac{\Delta H}{\Delta S_{\mathrm{mix}}}\right)
-20000\,|\delta|.}
\label{eq:sr_model}
\end{equation}
\rev{When values are reported in degrees Celsius, the conversion
$T_{\mathrm{SR}}(^\circ\mathrm{C})=T_{\mathrm{SR}}(\mathrm{K})-273.15$
is applied after evaluating Eq.~\ref{eq:sr_model}.}
\rev{Here, $\Delta H$ is the positive magnitude of the B2 formation enthalpy defined in Section~2.4,
$\Delta S_{\mathrm{mix}}=-k_{\mathrm B}\sum_i c_i\ln c_i$ is the ideal configurational entropy of the
corresponding disordered BCC matrix, and $\delta$ is the coherent lattice misfit strain. The first
term therefore defines the calibrated thermodynamic temperature scale, while the second term is
the fitted coherency penalty.}

\begin{figure}[H]
    \centering
    \includegraphics[width=\textwidth]{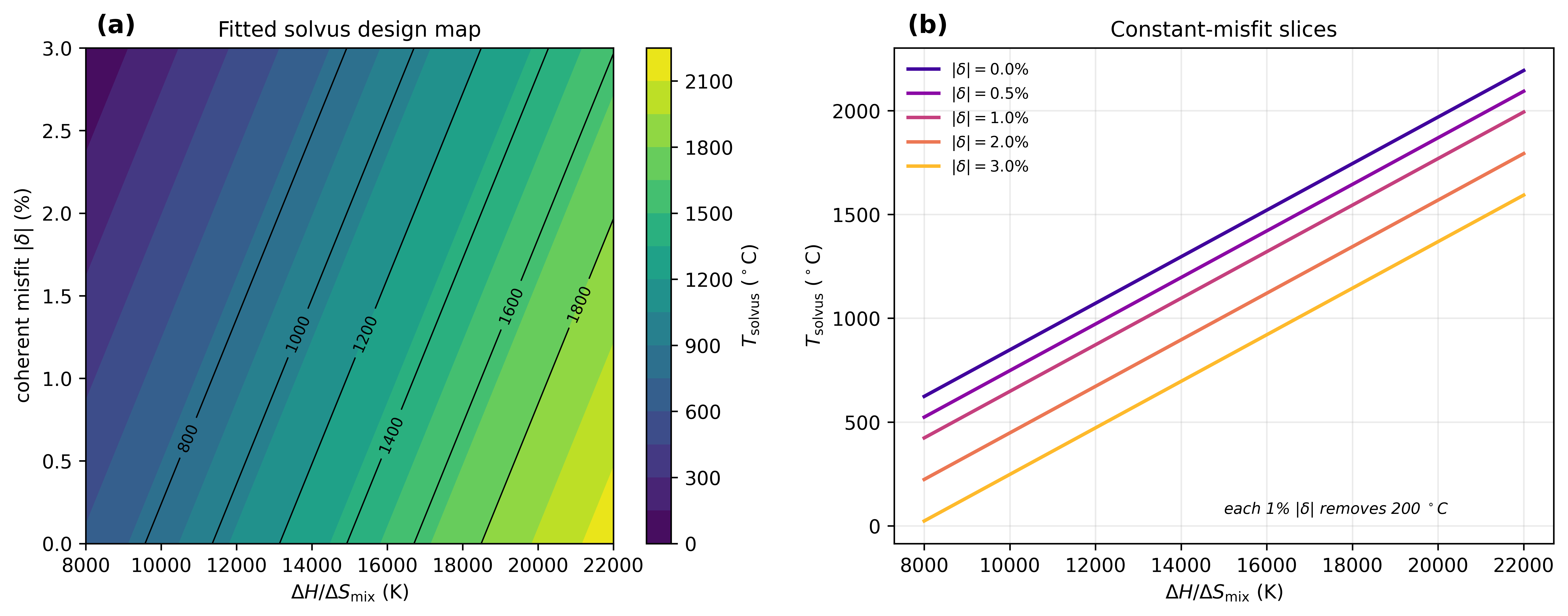}
    \caption{\rev{Symbolic-regression-based visualization of the fitted solvus relation. (a) Joint dependence on thermodynamic scale and lattice misfit. (b) Constant-misfit slices. The linear bands and parallel slices follow directly from Eq.~\ref{eq:sr_model} and are shown as a visualization of the fitted relation rather than as independent physical findings. The model-selection evidence for retaining the linear $|\delta|$ penalty over higher-order alternatives is reported separately in Supplementary Table~S8.}}
    \label{fig:sr_optimization}
\end{figure}

\rev{The structure of Eq.~\ref{eq:sr_model} is illustrated in Fig.~\ref{fig:sr_optimization}. Because
Eq.~\ref{eq:sr_model} is linear in $\Delta H/\Delta S_{\mathrm{mix}}$ and $|\delta|$, the
iso-stability contours in Fig.~\ref{fig:sr_optimization}(a) and the parallel constant-misfit slices
in Fig.~\ref{fig:sr_optimization}(b) are direct algebraic consequences of the fitted expression.
These graphical features are therefore used to visualize the model and are not interpreted as
independent physical findings. The quantitative result illustrated by the figure is the magnitude
of the fitted coherency penalty: each additional 1\% of $|\delta|$ lowers the effective solvus
prediction by 200~$^\circ$C. The model-selection evidence for retaining the linear $|\delta|$
dependence over the higher-order alternatives available to the symbolic-regression search is
reported separately in Supplementary Table~S8}.

Although classical continuum elasticity predicts that coherent strain energy scales quadratically
with misfit, expressions containing $\delta^2$ or higher-order misfit terms did not yield
statistically meaningful improvement during symbolic regression and were systematically
eliminated in favor of the linear $|\delta|$ dependence. This outcome reflects a scale-separation
and identifiability effect intrinsic to refractory BCC--B2 systems. In the present dataset, lattice
misfits typically lie in the range $|\delta| \approx 0.005$--0.03, such that squared misfit terms
contribute corrections of order $10^{-4}$--$10^{-3}$. Even when multiplied by large numerical
prefactors, these quadratic contributions correspond to solvus shifts of only a few tens of degrees,
which are small relative to the refractory solvus temperature scale of interest
($\sim$1000--2500~$^\circ$C) and comparable to experimental uncertainty, compositional scatter,
and calibration noise. As a result, they are not robustly identifiable on the macroscopic solvus
temperature scale.

The linear $|\delta|$ term should therefore be interpreted as an effective macroscopic descriptor
of coherency-limited stability rather than as a microscopic elastic energy law. After
coarse-graining over realistic precipitate size distributions, morphology transitions, and
progressive coherency loss, solvus suppression reflects the accumulation of elastic incompatibility
rather than the idealized elastic energy of a perfectly coherent inclusion. The symbolic regression
thus identifies the simplest physically meaningful form that remains predictive across the
experimentally relevant design space. \rev{The magnitude of the two terms is illustrated quantitatively in the worked example below. For the fixed (Ru$_{0.75}$Al$_{0.25}$)--Hf precipitate, $|E_{\mathrm{form}}^{\mathrm{B2}}|=0.926$~eV/atom. Against a binary 25/75 BCC matrix, $\Delta S_{\mathrm{mix}}=4.846\times10^{-5}$~eV~atom$^{-1}$~K$^{-1}$, giving $\Delta H/\Delta S_{\mathrm{mix}}\approx1.911\times10^{4}$~K and a calibrated thermodynamic contribution of approximately 1869~$^\circ$C. A 1\% misfit would then reduce the effective model prediction by 200~$^\circ$C. Table~\ref{tab:worked_example} applies the same arithmetic to four actual BCC matrices from the candidate dataset.}

\subsubsection{\rev{Separating the two terms of the solvus relation}}

\rev{To illustrate the separate thermodynamic and lattice-misfit contributions in Eq.~\ref{eq:sr_model}, Table~\ref{tab:worked_example} applies the relation to a fixed B2 precipitate, the Al-substituted Ru--Hf phase (Ru$_{0.75}$Al$_{0.25}$)--Hf, paired with four different BCC matrices. The relaxed B2 structure has $E_{\mathrm{form}}^{\mathrm{B2}}=-0.926$~eV/atom. Because all four matrices have binary 25/75 compositions, they have the same ideal configurational entropy, $\Delta S_{\mathrm{mix}}=4.846\times10^{-5}$~eV~atom$^{-1}$~K$^{-1}$. Using the magnitude of the B2 formation energy gives the same thermodynamic ratio, $\Delta H/\Delta S_{\mathrm{mix}}=19109$~K, for each pair. The thermodynamic contribution is therefore fixed at 1869~$^\circ$C, while the matrix-dependent lattice misfit produces penalties ranging from 2 to 330~$^\circ$C. This controlled comparison isolates the effect of the misfit term and shows directly how increasing $|\delta|$ lowers the predicted solvus even when the chemical driving force is unchanged. Additional explicit arithmetic verification of Eq.~\ref{eq:sr_model} for three representative candidates is provided in Supplementary Section~S5.}

{\color{red}
\begin{table}[H]
\centering
\caption{Worked application of Eq.~\ref{eq:sr_model} for one fixed B2 precipitate across several BCC matrices.}
\label{tab:worked_example}
\resizebox{\textwidth}{!}{%
\begin{tabular}{l l c c c c c}
\toprule
BCC matrix & Fixed B2 precipitate & $\Delta H/\Delta S_{\mathrm{mix}}$ (K) & $|\delta|$ & Thermodynamic term ($^\circ$C) & Misfit penalty ($^\circ$C) & $T_{\mathrm{SR}}$ ($^\circ$C) \\
\midrule
Mo$_{25}$Ta$_{75}$ & (Ru$_{0.75}$Al$_{0.25}$)--Hf & 19109 & 0.0001 & 1869 & 2 & 1867 \\
Nb$_{75}$Mo$_{25}$ & (Ru$_{0.75}$Al$_{0.25}$)--Hf & 19109 & 0.0032 & 1869 & 64 & 1805 \\
Nb$_{25}$Ta$_{75}$ & (Ru$_{0.75}$Al$_{0.25}$)--Hf & 19109 & 0.0143 & 1869 & 286 & 1583 \\
Nb$_{75}$Ta$_{25}$ & (Ru$_{0.75}$Al$_{0.25}$)--Hf & 19109 & 0.0165 & 1869 & 330 & 1539 \\
\bottomrule
\end{tabular}}
\end{table}
}

\subsubsection{Comparison to Experimental Solvus Temperatures}

\rev{Published alloys provide an external benchmark for the effective solvus relation. Table~\ref{tab:exp_solvus_comparison} maps each literature alloy to a chemistry-matched DFT proxy using the stoichiometric RuHf, RuZr, and RuTi binaries. The experimental solvus temperatures and lattice misfits are taken from the sequential-annealing and diffraction measurements of Frey and co-workers on HfRu- and ZrRu-B2 alloys in Nb--V matrices \cite{Frey2024b} and from the broader survey of Ru-B2 alloys by Kube and co-workers \cite{Kube2024}. For each comparison, the thermodynamic contribution is evaluated using the DFT formation enthalpy of the corresponding binary B2 phase and the ideal configurational entropy of the (Nb,V) or (Nb,Mo) matrix, while the misfit penalty is evaluated using the measured room-temperature lattice misfit.}

\rev{The comparison shows a systematic pattern. The thermodynamic term alone gives effective temperatures of approximately 1800--2340~$^\circ$C for the near-stoichiometric binaries, consistent with the strong intrinsic stability of RuHf, RuZr, and RuTi. The resulting relation therefore represents an effective upper-bound design temperature for the near-stoichiometric B2 phase rather than a directly calculated equilibrium solvus. The experimental alloys contain substantially lower Ru concentrations, typically 2--9~at.\%, and consequently exhibit lower effective solvus temperatures that increase with Ru content. For the Hf-containing alloys, the reported solvus increases from below 1000~$^\circ$C at approximately 2~at.\% Ru, to 1300--1400~$^\circ$C at approximately 4~at.\% Ru \revB{(Hf$_5$Ru$_4$--Nb$_{63}$V$_{28}$)}, and to about 1750~$^\circ$C at approximately 9~at.\% Ru \revB{(Hf$_{11}$Ru$_9$--Nb$_{52}$V$_{28}$). At the same Ru content of about 9~at.\%, the Zr-containing alloys Zr$_{10}$Ru$_9$--Nb$_{63}$V$_{18}$ and Hf$_2$Zr$_8$Ru$_9$--Nb$_{63}$V$_{18}$ show lower solvus temperatures of 1300--1500~$^\circ$C, whereas the 1750--1900~$^\circ$C range in Table~\ref{tab:exp_solvus_comparison} corresponds to the Ti-containing Ru$_{15}$Ti$_{15}$--Nb$_{35}$Mo$_{35}$ alloy at 15~at.\% Ru.} The calculated relation reproduces the overall ordering and the family-level trend, including the greater stability of HfRu relative to ZrRu, while not being intended to reproduce the solvus of each dilute alloy exactly. Under Eq.~\ref{eq:sr_model}, $\partial T_{\mathrm{solvus}}/\partial|\delta|=-20000$~K for every chemical family. Ru--Zr therefore does not have a different intrinsic sensitivity to lattice misfit from Ru--Hf within the fitted model. Any lower predicted temperature for the Zr family arises instead from a lower thermodynamic contribution and/or a larger actual $|\delta|$ for the specific matrix--precipitate pair.}

\rev{The dilute Hf$_2$Ru$_1$--Nb$_{69}$V$_{28}$ alloy is considered separately. At only 2~at.\% Hf and 1~at.\% Ru, it lies outside the near-stoichiometric B2 domain represented in the DFT library, and Eq.~\ref{eq:sr_model} is therefore not applied to this composition. The reported single-phase BCC state is consistent with the dilute Hf and Ru additions remaining accommodated within the BCC solid solution rather than forming a distinct B2 precipitate.}

{\color{red}
\begin{table}[H]
\centering
\caption{\rev{Literature benchmark of the symbolic-regression relation using chemistry-matched near-stoichiometric B2 proxies. $T_{\mathrm{SR}}$ is the predicted effective design temperature, while $T_{\mathrm{exp}}$ denotes the reported experimental solvus range \cite{Frey2024b,Kube2024}.}}
\label{tab:exp_solvus_comparison}
\resizebox{\textwidth}{!}{%
\begin{tabular}{l l l c c c c c c}
\toprule
Literature alloy & BCC proxy & B2 proxy & $\Delta H/\Delta S$ (K) & $|\delta|_{\mathrm{exp}}$ & Thermo. ($^\circ$C) & Penalty ($^\circ$C) & $T_{\mathrm{SR}}$ ($^\circ$C) & $T_{\mathrm{exp}}$ ($^\circ$C) \\
\midrule
Hf$_{11}$Ru$_9$--Nb$_{52}$V$_{28}$ & (Nb,V) & RuHf & 22240 & 0.5\% & 2220 & 100 & 2120 & $\sim$1750 \\
Hf$_5$Ru$_4$--Nb$_{63}$V$_{28}$ & (Nb,V) & RuHf & 23320 & 0.5\% & 2341 & 100 & 2241 & 1300--1400 \\
Zr$_{10}$Ru$_9$--Nb$_{63}$V$_{18}$ & (Nb,V) & RuZr & 22250 & 0.5\% & 2221 & 100 & 2121 & 1300--1400 \\
Hf$_2$Zr$_8$Ru$_9$--Nb$_{63}$V$_{18}$ & (Nb,V) & (Hf,Zr)Ru & 23240 & 0.8\% & 2332 & 160 & 2172 & 1400--1500 \\
Ru$_{15}$Ti$_{15}$--Nb$_{35}$Mo$_{35}$ & (Nb,Mo) & RuTi & 18720 & 3.6\% & 1825 & 720 & 1105 & 1750--1900 \\
Hf$_3$Ru$_2$--Nb$_{67}$V$_{28}$ (dilute) & -- & -- & -- & -- & -- & -- & outside domain & $<1000$ \\
\bottomrule
\end{tabular}}
\end{table}
}

\rev{Two rows require comment. The Zr$_{10}$Ru$_9$ and Hf$_5$Ru$_4$ alloys have almost equal effective solvus of 1300 to 1400~$^\circ$C despite the higher ruthenium content of the zirconium alloy, which the measurements of Frey and co-workers attribute to the lower thermodynamic driving force of ZrRu and to tertiary-phase formation that removes ruthenium from the matrix. This is consistent with Eq.~\ref{eq:sr_model}, in which the zirconium thermodynamic term is lower than the hafnium term. The Ru$_{15}$Ti$_{15}$ alloy shows the opposite behaviour, with a predicted temperature that falls well below the measured range once the large room-temperature misfit of 3.6~percent is applied. This is expected, because the misfit that controls coherency at the annealing temperature is smaller than the room-temperature value owing to the different thermal expansion of the two phases, an effect noted directly in the source measurements. When the coherency-relevant misfit at temperature is used instead, the thermodynamic term of about 1825~$^\circ$C is consistent with the measured 1750 to 1900~$^\circ$C. The benchmark therefore supports the structure of the relation and its ranking of families, while making explicit that the room-temperature misfit is an upper bound on the penalty and that the effective solvus of a dilute alloy sits below the intrinsic prediction.}


\subsection{Elastic Penalties and Compositional Selection}
\label{sec:elastic_penalties}

\begin{figure}[H]
    \centering

    \includegraphics[width=0.80\textwidth]{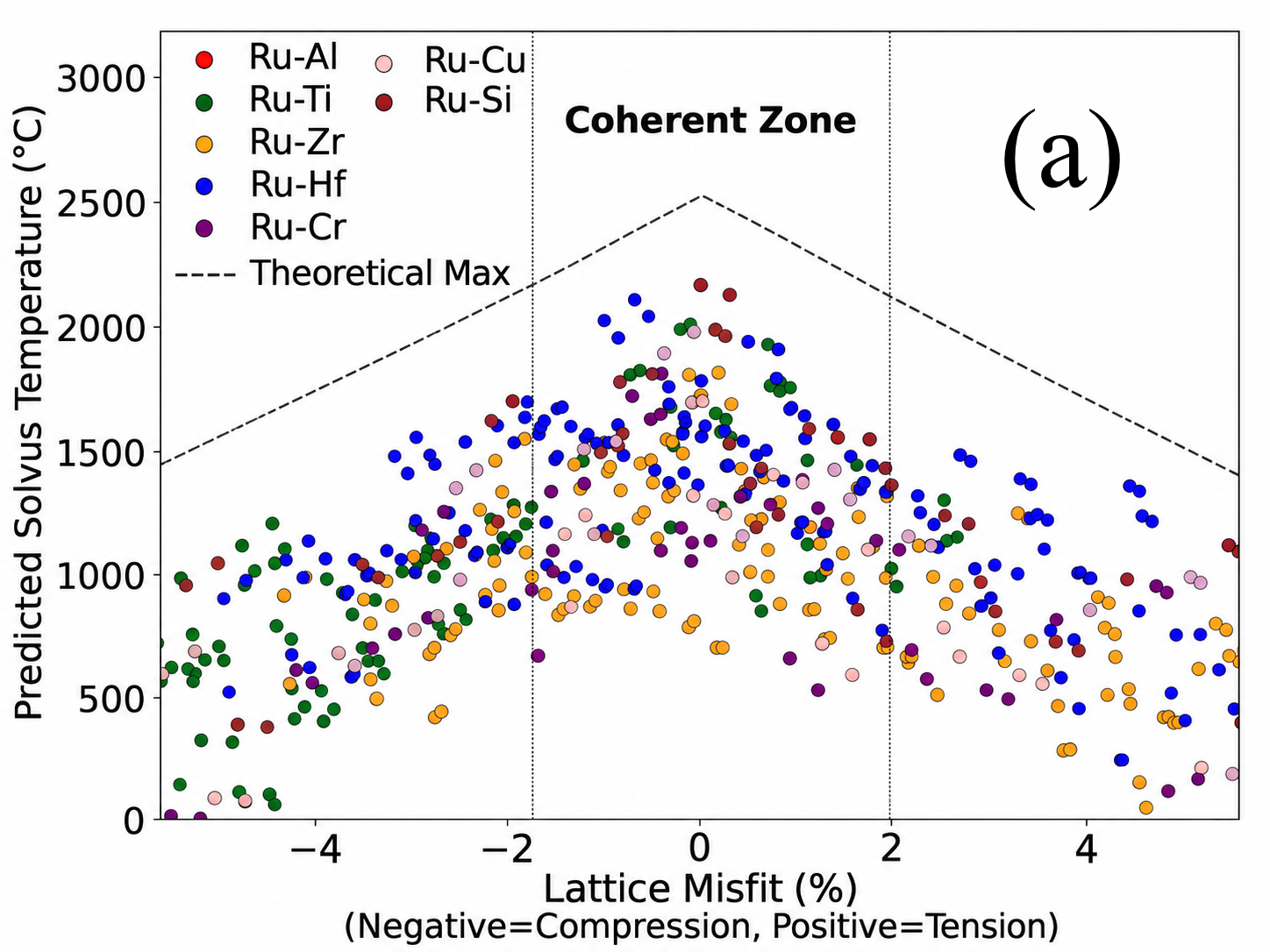}

    \vspace{0.3cm}

    \includegraphics[width=0.80\textwidth]{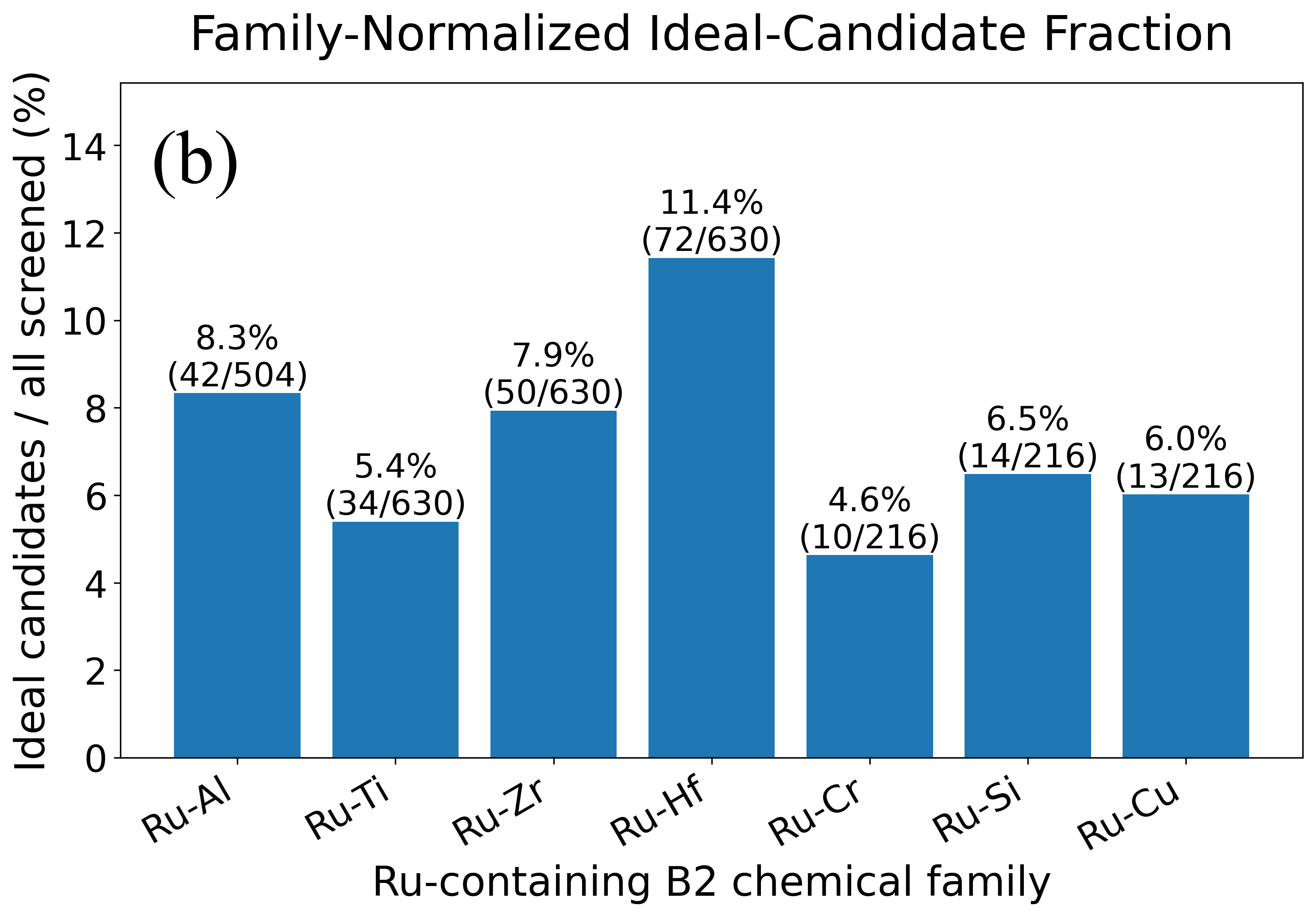}

    \caption{\rev{Composition-selection summary. 
    (a) Predicted solvus temperature as a function of lattice misfit for the sampled Ru-containing B2 chemical families. The dotted vertical lines indicate the $\pm2\%$ coherency window, while the dashed envelope indicates the maximum temperature predicted by the fitted symbolic-regression relation at each misfit. 
    (b) Family-normalized ideal-candidate fraction, defined as the number of ideal candidates in each Ru--$X$-containing B2 family divided by the total number screened in that family, $N_{\mathrm{ideal}}/N_{\mathrm{screened}}$. The normalization removes bias associated with unequal sampling across chemical families. Multicomponent candidates may contribute to more than one Ru--$X$ family when the corresponding element is present in the B2 composition.}}
    
    \label{fig:elemental_mechanism}
\end{figure}

\rev{The fitted $|\delta|$ term is interpreted as an effective macroscopic descriptor of coherency-limited stability over the sampled composition space rather than as a microscopic elastic-energy law. Classical coherent strain energy scales quadratically with lattice misfit, whereas the symbolic-regression model identifies a linear absolute-misfit penalty over the present design range. Figure~\ref{fig:elemental_mechanism}(a) provides a direct visualization of this fitted dependence. The predicted solvus temperature is highest near $\delta=0$ and decreases symmetrically as $|\delta|$ increases, while the dotted lines at $\pm2\%$ indicate a representative coherency window. The dashed envelope represents the maximum temperature predicted by the fitted relation at each misfit. The functional form itself is established by the symbolic-regression model selection summarized in Supplementary Table~S8 rather than by the geometry of the plotted envelope. The candidate ranking is therefore interpreted on the basis of individual B2--BCC matrix pairings rather than by assigning a universal coherency advantage to an entire binary family. Stoichiometric RuHf and RuZr can differ from single-element refractory BCC matrices by several percent in lattice parameter, whereas selected off-stoichiometric or multicomponent members of the same chemical families can approach the low-misfit region. The practical design objective is therefore to preserve a favorable thermodynamic contribution while tuning the particular precipitate--matrix combination toward $|\delta|\simeq0$.}

\rev{Figure~\ref{fig:elemental_mechanism}(b) quantifies the chemical-family dependence using a normalized success fraction, defined as $N_{\mathrm{ideal}}/N_{\mathrm{screened}}$ for each Ru--$X$-containing B2 family. This normalization is necessary because the different chemical families are not sampled equally in the candidate library. Within the present screening space, Ru--Hf-containing candidates have the highest ideal-candidate fraction, $72/630=11.4\%$, followed by Ru--Al ($42/504=8.3\%$) and Ru--Zr ($50/630=7.9\%$). The corresponding fractions are $6.5\%$ for Ru--Si, $6.0\%$ for Ru--Cu, $5.4\%$ for Ru--Ti, and $4.6\%$ for Ru--Cr. These values should be interpreted as relative enrichment within the sampled design space rather than as universal probabilities of B2 formation. For multicomponent B2 compositions, a candidate may contribute to more than one Ru--$X$ chemical family when multiple alloying elements are present.}

\subsubsection{Lattice Tuning and the Multi-Component Stabilization of Ru-B2 Precipitates}

\rev{A useful feature of the candidate list is the limited representation of stoichiometric binary B2 compounds such as RuHf, RuTi, or RuZr despite their strong intrinsic ordering. Stoichiometric binaries paired with single-element refractory BCC matrices can exhibit lattice mismatches of a few percent, whereas selected off-stoichiometric and multicomponent members of the same chemical families enter the near-zero-misfit window. This distinction emphasizes that coherency must be evaluated for the specific precipitate--matrix pair using the relaxed lattice parameters rather than inferred from the binary chemical identity alone. Consequently, multicomponent lattice engineering provides a practical route for tuning otherwise favorable B2 chemistries. The ideal candidates identified by the model are frequently ternary and quaternary compositions (e.g., Ru--Hf--Al and Ru--Ti--Zr) in which the B2 sublattice balances elements of different atomic size. For instance, Frey et al. \cite{Frey2024b} report higher thermal stability for HfRu than for ZrRu, but realizing this thermodynamic advantage in a specific BCC matrix also requires an acceptable lattice match. Partial replacement of Ru by Al or related additions changes the relaxed B2 lattice parameter and can move a selected Ru--Hf/BCC pair toward $\delta\rightarrow0$. This reduces the fitted $20000|\delta|$ penalty, while the concentration-dependent analysis below quantifies the accompanying thermodynamic cost. Si- and Cr-containing candidates show less favorable combined thermodynamic/misfit scores in parts of the screened design space. No separate electronic mechanism is assigned to those trends from the screening model alone. The targeted electronic-structure calculations in Section~3.8 are used only for the representative chemistries explicitly analyzed there. The general design requirement is therefore that a useful substitution must improve the lattice match faster than it degrades the thermodynamic contribution.}


\subsubsection{\rev{Quantifying the Al and Si tuning cross-over from the existing substitution series}}

\rev{The DFT dataset contains the A-sublattice series $(\mathrm{Ru}_{1-x}M_x)$--Hf at Ru fractions 1.0, 0.875, 0.75, 0.625, and 0.5 for $M=$Al and Si. The relaxed values are tabulated in the Supplementary Material. Over the sampled range, the aluminium series raises the B2 lattice parameter from 3.239 to 3.344~\AA\ and lowers the magnitude of the formation enthalpy from 1.19 to 0.73~eV/atom, so both the misfit and the thermodynamic contribution change systematically with composition. Against the representative Nb$_{75}$V$_{25}$ matrix, the misfit passes through zero near $x^{*}\simeq0.11$, corresponding to approximately 11~at.\% aluminium, while the silicon series crosses zero slightly earlier near $x^{*}\simeq0.07$.}

\rev{Figure~\ref{fig:al_si_tuning} shows that, for the Ru--Hf system, the thermodynamic cost of substitution outweighs the coherency benefit as the substitution level increases. Each 1~percent reduction in misfit removes approximately 200~$^\circ$C of the fitted coherency penalty, whereas increasing $x$ from 0 to 0.5 lowers the thermodynamic contribution by roughly 1000~$^\circ$C. Consequently, the effective solvus decreases monotonically with substitution rather than passing through a maximum. Aluminium and silicon therefore act as lattice-tuning additions that improve coherency at a thermodynamic cost. For an initially mismatched B2--BCC pair, the preferred strategy is to use the smallest substitution required to enter the coherent window, near $x^{*}$, rather than to continue increasing the substitution level beyond the lattice-matching condition.}

{\color{red}
\begin{figure}[H]
\centering
\includegraphics[width=\textwidth]{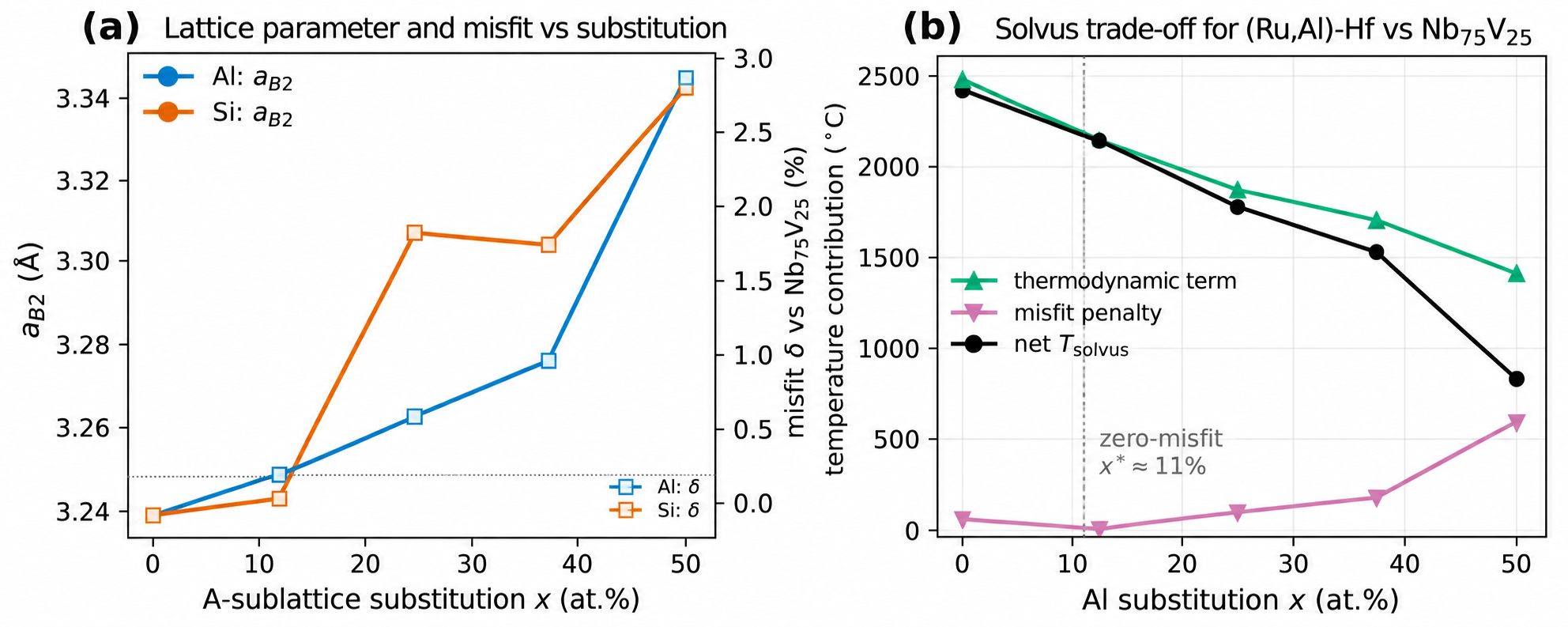}
\caption{\rev{Concentration-dependent lattice tuning in $(\mathrm{Ru}_{1-x}M_x)$--Hf for $M=$ Al and Si. (a) Relaxed B2 lattice parameter and lattice misfit relative to Nb$_{75}$V$_{25}$ as a function of A-sublattice substitution. (b) Thermodynamic contribution, misfit penalty, and net effective solvus for the (Ru,Al)--Hf series.}}
\label{fig:al_si_tuning}
\end{figure}
}

\subsection{\rev{Independent Thermodynamic and Structural Benchmarks}}
\label{sec:independent_benchmark}

\begin{figure}[H]
\centering
\includegraphics[width=\textwidth]{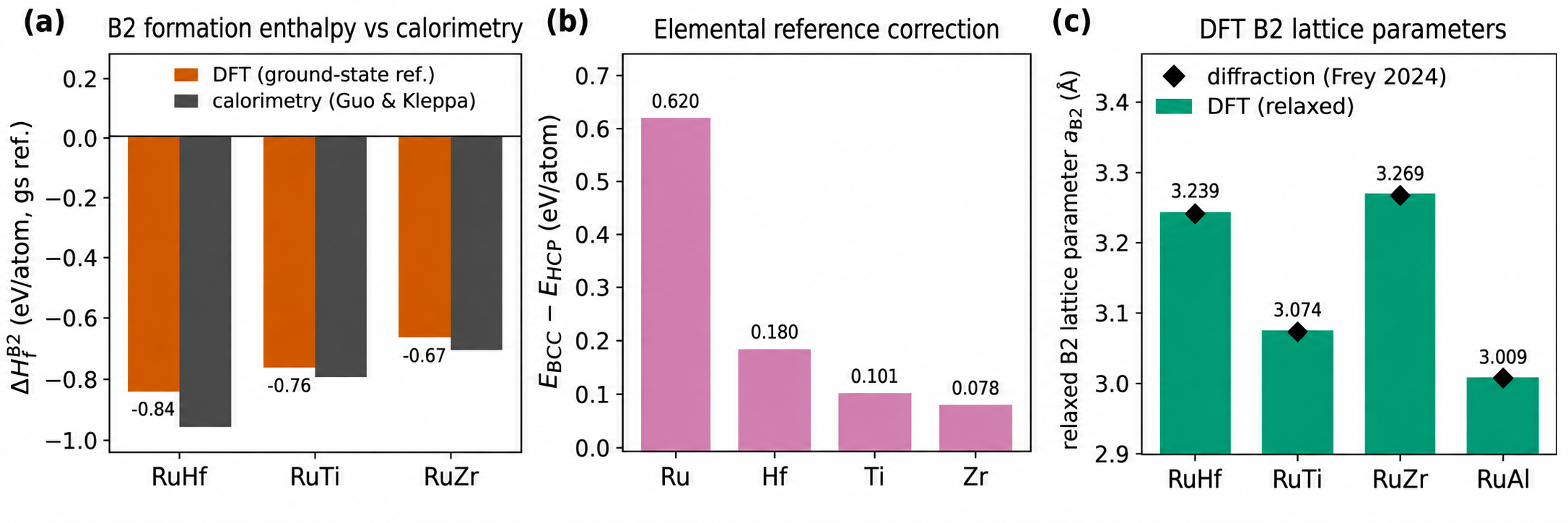}
\caption{\rev{Independent benchmarking of the calculated thermodynamic and structural quantities. (a) Ground-state-referenced DFT formation enthalpies of RuHf, RuTi, and RuZr compared with high-temperature direct-synthesis calorimetry. (b) Matched-settings elemental reference correction $E_{\mathrm{BCC}}-E_{\mathrm{HCP}}$ for Ru, Hf, Ti, and Zr. (c) Relaxed DFT B2 lattice parameters compared with published diffraction values for RuHf, RuTi, RuZr, and RuAl.}}
\label{fig:independent_benchmark}
\end{figure}

\rev{Independent published measurements are used to test the thermodynamic and structural quantities that enter the computational framework. Table~\ref{tab:independent_benchmark} summarizes these comparisons on a common basis. The thermodynamic benchmark uses high-temperature direct-synthesis calorimetry for RuHf, RuTi, RuZr, and RuAl \cite{GuoKleppa2001,SuNash2005}. The calculated formation enthalpies are placed on the elemental ground-state reference scale: RuHf, RuTi, and RuZr use the matched HCP/BCC reference corrections listed in Table~\ref{tab:reference_state_revision}, while RuAl uses the FCC ground state of aluminium. The structural benchmark compares the relaxed B2 lattice parameters with diffraction measurements compiled by Frey and co-workers \cite{Frey2024b}. The table also lists the refractory BCC lattice parameters relevant to the matrix side of the misfit calculation and the near-linear Ru--X B2 composition--lattice-parameter trend reported by Kube and co-workers \cite{Kube2024}. Figure~\ref{fig:independent_benchmark} separates these checks into their individual components. Figure~\ref{fig:independent_benchmark}(a) compares the ground-state-referenced DFT formation enthalpies of RuHf, RuTi, and RuZr directly with calorimetry. The calculated values are $-81.2$, $-73.1$, and $-64.3$~kJ~mol$^{-1}$-atom, respectively, compared with experimental values of $-91.8$, $-77.0$, and $-68.7$~kJ~mol$^{-1}$-atom. The calculations reproduce both the sign and the chemical ordering of the formation enthalpies, with RuHf being the most strongly bound and RuZr the least strongly bound of these three systems. Table~\ref{tab:independent_benchmark} additionally shows that the ground-state-referenced RuAl formation enthalpy is $-62.2$~kJ~mol$^{-1}$-atom, compared with the calorimetric value of $-54.5$~kJ~mol$^{-1}$-atom. The largest deviation among the four thermochemical benchmarks is approximately 11~kJ~mol$^{-1}$-atom.}

\rev{Figure~\ref{fig:independent_benchmark}(b) shows the elemental reference-state corrections used to place these quantities on the appropriate ground-state scale. The calculated $E_{\mathrm{BCC}}-E_{\mathrm{HCP}}$ values are 0.620, 0.180, 0.101, and 0.078~eV/atom for Ru, Hf, Ti, and Zr, respectively. These corrections are particularly important for Ru because the BCC reference lies substantially above the HCP ground state, and they prevent the absolute B2 formation enthalpies from being compared with calorimetry on inconsistent elemental reference scales. As discussed in Section 2.1.1, these reference shifts alter the absolute formation enthalpy but cancel for composition-matched B2/BCC energy differences. Figure~\ref{fig:independent_benchmark}(c) provides the corresponding structural benchmark. The relaxed DFT B2 lattice parameters are 3.239, 3.074, 3.269, and 3.009~\AA\ for RuHf, RuTi, RuZr, and RuAl, respectively. These values closely reproduce the reported diffraction values: 3.225~\AA\ for RuHf, 3.060--3.070~\AA\ for RuTi, 3.253~\AA\ for RuZr, and 2.992--3.036~\AA\ for RuAl. The deviations for RuHf, RuTi, and RuZr are approximately 0.4--0.5\%, while the RuAl calculation lies within the reported experimental range. This level of agreement is important because the coherency design window considered in this work is itself of order one percent.}

{\color{red}
\begin{table}[H]
\centering
\small
\caption{\rev{Independent thermodynamic and structural quantities used for benchmarking. Formation enthalpies are referenced to the elemental ground states, while the structural quantities are compared with published diffraction data. Experimental thermochemical and structural values are taken from the cited literature.}}
\label{tab:independent_benchmark}

\setlength{\tabcolsep}{4pt}
\renewcommand{\arraystretch}{1.08}

\resizebox{\textwidth}{!}{%
\begin{tabular}{l l c c c l}
\toprule
Quantity & Phase/system & Experiment & DFT value & Difference & Source \\
\midrule
Formation enthalpy & RuHf & $-91.8$~kJ/mol-at & $-81.2$~kJ/mol-at & $+10.6$ & \cite{GuoKleppa2001,Frey2024b} \\
Formation enthalpy & RuTi & $-77.0$~kJ/mol-at & $-73.1$~kJ/mol-at & $+3.9$ & \cite{GuoKleppa2001,Frey2024b} \\
Formation enthalpy & RuZr & $-68.7$~kJ/mol-at & $-64.3$~kJ/mol-at & $+4.4$ & \cite{GuoKleppa2001,Frey2024b} \\
Formation enthalpy & RuAl & $-54.5$~kJ/mol-at & $-62.2$~kJ/mol-at & $-7.7$ & \cite{SuNash2005,Frey2024b} \\

Lattice parameter & RuHf & $3.225$~\AA & $3.239$~\AA & $+0.4\%$ & \cite{Frey2024b} \\
Lattice parameter & RuTi & $3.060$--$3.070$~\AA & $3.074$~\AA & $+0.3\%$ & \cite{Frey2024b} \\
Lattice parameter & RuZr & $3.253$~\AA & $3.269$~\AA & $+0.5\%$ & \cite{Frey2024b} \\
Lattice parameter & RuAl & $2.992$--$3.036$~\AA & $3.009$~\AA & $<0.3\%$ & \cite{Frey2024b} \\

Lattice parameter (BCC) &
\shortstack[l]{Nb/Mo/Ta/\\W/V} &
\shortstack{$3.301/3.147/3.303/$\\$3.165/3.026$~\AA} &
\shortstack{$3.328/3.154/3.300/$\\$3.171/2.972$~\AA} &
\shortstack{$+0.8/{+}0.2/{-}0.1/$\\${+}0.2/{-}1.8\%$} &
\cite{Frey2024b} \\

Pseudo-binary trend & Ru--(Al,Ti,Hf,Zr) B2 &
near-linear (Vegard) & near-linear & slope agrees & \cite{Kube2024} \\
\bottomrule
\end{tabular}%
}
\end{table}
}

\subsection{Transferability to Higher-Order Refractory Matrices and Domain of Validity}
\label{sec:transferability_v2}

\rev{Having established that the thermodynamic and structural quantities underlying the design framework are consistent with independent published measurements, we next examine whether the same lattice-matching criterion remains applicable as the refractory BCC matrix becomes compositionally more complex. This provides a bounded test of transferability beyond the binary and ternary matrices that constitute much of the lower-order screening space, without changing the Ru-containing B2 chemistry used to construct the design framework. The matrix library spans all binary combinations of Nb, Mo, Ta, W, and V over several compositions, all ten equiatomic ternary BCC matrices formed from these elements, five equiatomic quaternary BCC SQS matrices, and the equiatomic quinary NbMoTaVW matrix. The higher-order matrices are paired with the same B2 library so that the effect of increasing matrix complexity can be examined independently of changes in precipitate chemistry. The relaxed BCC-equivalent lattice parameters of NbMoTaW, NbMoTaV, NbTaVW, MoTaVW, NbMoVW, and NbMoTaVW are 3.2315, 3.1956, 3.2008, 3.1534, 3.1583, and 3.1887~\AA, respectively.}

\rev{Figure~\ref{fig:multicomponent_generalization_v2} summarizes this higher-order matrix test. Figure~\ref{fig:multicomponent_generalization_v2}(a) compares the relaxed lattice parameters of the quaternary and quinary matrices with representative RuHf, RuTi, RuZr, and RuAl B2 lattice parameters. The higher-order matrices span a sufficiently broad lattice-parameter range to generate both closely matched and strongly mismatched matrix--precipitate combinations. This is quantified in Fig.~\ref{fig:multicomponent_generalization_v2}(b), which shows the corresponding coherent lattice misfit for each B2/matrix pair. For example, NbMoTaW is nearly lattice matched to RuHf, with a calculated misfit of approximately $+0.2\%$, whereas the same matrix is strongly mismatched with RuTi and RuAl. Similar pair-specific behavior is observed across the other higher-order matrices. Thus, increasing matrix compositional complexity does not remove the basic geometric constraint identified in the lower-order systems; rather, it expands the range of lattice parameters available for identifying favorable B2--BCC combinations. The transferability of the framework is therefore deliberately bounded to Ru-containing B2 precipitates in Nb--Mo--Ta--W--V-based refractory BCC matrices over the sampled composition and misfit ranges. It is not intended to describe arbitrary dilute solid solutions, unrelated matrix chemistries, or regimes in which competing Laves, A15, $\sigma$, or silicide phases control phase stability. Likewise, the design map ranks candidate matrix--precipitate combinations according to thermodynamic and coherency criteria and should not be interpreted as a predictor of complete mechanical performance.}

{\color{red}
\begin{figure}[H]
\centering
\includegraphics[width=\textwidth]{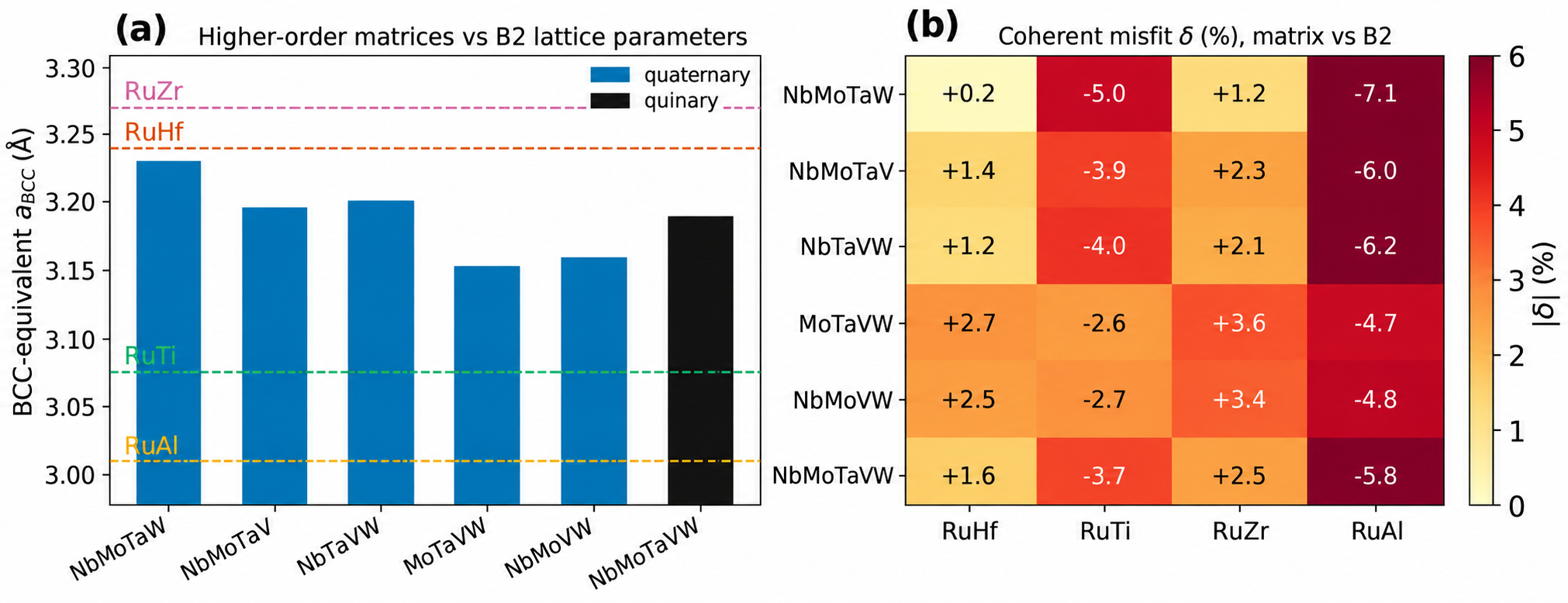}
\caption{\rev{Transferability to higher-order refractory matrices. (a) Relaxed BCC-equivalent lattice parameters of five equiatomic quaternary matrices and equiatomic NbMoTaVW, with representative Ru-based B2 lattice parameters shown as reference lines. (b) Coherent lattice misfit of each higher-order matrix against representative RuHf, RuTi, RuZr, and RuAl B2 phases.}}
\label{fig:multicomponent_generalization_v2}
\end{figure}
}

\subsection{Targeted Electronic-Structure Interpretation}
\label{sec:electronic_v2}

\rev{The differences in formation energy and lattice parameter among the Ru-based B2 phases are accompanied by systematic differences in their electronic structure. To examine these trends, static projected-density-of-states calculations were performed for five representative relaxed structures from the DFT dataset: RuHf, RuTi, RuZr, RuAl, and $(\mathrm{Ru}_{0.75}\mathrm{Al}_{0.25})$--Hf. Tetrahedron integration and orbital projections were used to characterize the Ru $d$ states, the contributions from the partner elements, the density of states at the Fermi level, $N(E_F)$, and the Ru $d$-band centre. Figures~\ref{fig:electronic_v2}(a)--(c) show the projected densities of states of RuHf, RuTi, and RuZr. All three Ru--early-transition-metal B2 phases exhibit a pronounced pseudogap near the Fermi level. Their total $N(E_F)$ values are approximately 0.22, 0.26, and 0.30~states/eV/atom for RuHf, RuTi, and RuZr, respectively. In each case, the occupied states below $E_F$ contain strong contributions from both Ru $d$ states and the $d$ states of Hf, Ti, or Zr. The Ru $d$-band centres range from approximately $-1.26$ to $-1.42$~eV relative to $E_F$, indicating broadly similar electronic characteristics among these three B2 binaries.}

\rev{Figure~\ref{fig:electronic_v2}(d) shows that RuAl is electronically distinct from the three early-transition-metal binaries. Its density of states at the Fermi level increases to approximately 0.86~states/eV/atom, and the pseudogap is considerably less pronounced. The Al $p$ states overlap with the Ru $d$ states throughout part of the occupied-energy range, consistent with Ru--Al $p$--$d$ hybridization. The Ru $d$-band centre is also shifted toward the Fermi level relative to RuHf, RuTi, and RuZr. Figure~\ref{fig:electronic_v2}(e) shows the projected density of states of the Al-substituted Ru--Hf phase. The principal Hf-derived features remain present, while Al introduces additional $p$-state contributions near and below the Fermi level. Its $N(E_F)\approx0.58$~states/eV/atom lies between those of RuHf and RuAl, demonstrating that partial Al substitution modifies the electronic structure in addition to changing the lattice parameter.}

\rev{Figure~\ref{fig:electronic_v2}(f) compares the Ru $d$-band centres and $N(E_F)$ values of all five structures. RuHf, RuTi, and RuZr remain grouped at relatively low $N(E_F)$, whereas RuAl exhibits both a substantially larger $N(E_F)$ and a Ru $d$-band centre closer to $E_F$. The Al-substituted Ru--Hf structure has an intermediate $N(E_F)$, but its Ru $d$-band centre is deeper than those of the four binary compounds, showing that the two electronic descriptors do not vary monotonically with Al substitution. The electronic-structure results therefore indicate that the compositional trends among the Ru-based B2 phases cannot be described by atomic size and lattice matching alone. Changes in the Ru $d$-band position, pseudogap character, and Ru--Al $p$--$d$ interaction accompany the structural and thermodynamic changes produced by substitution. These electronic descriptors are used here as supporting interpretation rather than as independent predictors of phase stability. The analysis is restricted to the five representative structures for which projected electronic-structure calculations were performed.}

{\color{red}
\begin{figure}[H]
\centering
\includegraphics[width=\textwidth]{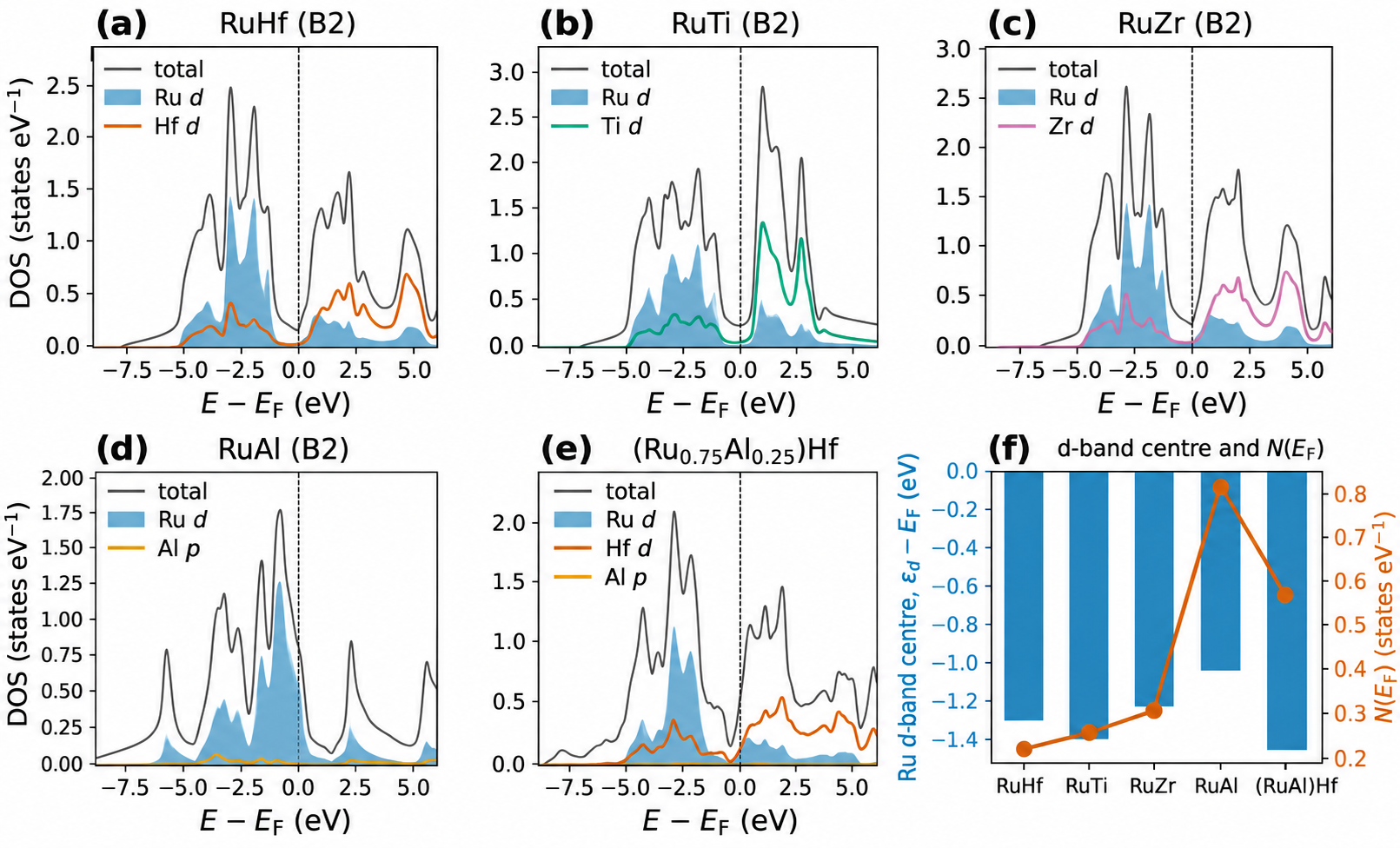}
\caption{\rev{Targeted electronic-structure comparison. Panels (a)--(e) show element- and orbital-projected densities of states for RuHf, RuTi, RuZr, RuAl, and $(\mathrm{Ru}_{0.75}\mathrm{Al}_{0.25})$--Hf, referenced to the Fermi level and normalized per atom. The dashed vertical line marks $E_F$. Panel (f) summarizes the Ru $d$-band centre and the total density of states at the Fermi level.}}
\label{fig:electronic_v2}
\end{figure}
}

\section{Conclusion}

Ru-containing B2 precipitates in refractory BCC matrices were investigated using high-throughput DFT calculations, machine-learning surrogate models, and symbolic regression. The DFT calculations provided formation energies and relaxed lattice parameters for evaluating the thermodynamic stability of the B2 phases and their lattice misfit with candidate BCC matrices. The calculations show that strongly ordered Ru-based B2 phases are not necessarily the most suitable coherent precipitates, because a favorable formation enthalpy can be offset by a large B2--BCC lattice mismatch. The random-forest model was used to extend the DFT trends across the candidate space, while symbolic regression produced the effective solvus relation
\[
T_{\mathrm{SR}}(\mathrm{K})
\approx
0.1121\,\frac{\Delta H}{\Delta S_{\mathrm{mix}}}
-
20000|\delta|.
\]
The relation shows explicitly that the predicted stability of a B2--BCC pair depends on both the thermodynamic ordering scale and the lattice mismatch.

\rev{The effect of chemical substitution was quantified using the Al- and Si-substituted Ru--Hf series. Both elements can tune the B2 lattice parameter toward the matrix value, but the improvement in coherency is accompanied by a reduction in the magnitude of the B2 formation enthalpy. For the representative Nb$_{75}$V$_{25}$ matrix, the zero-misfit condition occurs near 11~at.\% Al and 7~at.\% Si. Further substitution moves the lattice away from the matched condition while continuing to reduce the thermodynamic contribution. The preferred composition is therefore not the one with the largest amount of lattice-tuning addition, but the smallest substitution required to reach the desired coherency range. The calculated thermodynamic and structural quantities were compared with independent published measurements. Ground-state-referenced formation enthalpies of RuHf, RuTi, and RuZr reproduce the experimental chemical ordering and differ from calorimetric values by at most approximately 11~kJ~mol$^{-1}$-atom. The relaxed B2 lattice parameters of RuHf, RuTi, RuZr, and RuAl agree with reported diffraction values to within approximately 0.5\%. Published solvus measurements provide an additional external comparison for the effective temperature ranking. Differences between the static 0~K calculations and measured phase stability can arise from finite-temperature free-energy contributions, thermal expansion, competing phases, and kinetic effects that are not included explicitly in the present calculations.}

\rev{The same lattice-matching analysis was extended to five equiatomic quaternary BCC matrices and the equiatomic NbMoTaVW matrix. Near-coherent and strongly mismatched B2--BCC combinations are obtained within these higher-order matrices, showing that the lattice-matching criterion is not limited to binary and ternary matrix compositions. The electronic-structure calculations provide an additional interpretation of the chemical trends. RuHf, RuTi, and RuZr exhibit a pronounced pseudogap near the Fermi level and relatively low $N(E_F)$, whereas RuAl shows a substantially higher $N(E_F)$ and Ru $d$--Al $p$ hybridization. Partial Al substitution in Ru--Hf modifies this electronic structure while simultaneously changing the lattice parameter. Thus, lattice tuning changes both the geometric compatibility with the BCC matrix and the electronic contribution to B2 stability.}

\rev{The present model is restricted to Ru-containing B2 precipitates in Nb--Mo--Ta--W--V-based refractory BCC matrices over the sampled composition and misfit ranges. It is intended for ranking candidate matrix--precipitate combinations rather than predicting complete finite-temperature phase equilibria or mechanical properties. Experimental measurements on deliberately lattice-tuned Ru--Hf--Al and Ru--Zr--Al alloys would provide a direct test of the predicted composition--coherency trends.}


\section*{Acknowledgments}
This work was supported by the National Science Foundation through ACCESS supercomputing allocations MAT250103 and MAT240094. Additional computational resources were provided by Argonne National Laboratory under the Director’s Discretionary allocation for the project \textit{AIAlloyLW, GNNMD}. In addition to this, computational resources for this work were provided by a National Science Foundation MRI Award granted to Wilkes University (Award 1920129).

\section*{Author contribution}
Avik Mahata: Conceptualization, Methodology, Software, Formal analysis, Writing---Original draft preparation.

\section*{Data availability}
All datasets generated and analyzed during this study, including the high-throughput DFT formation energies and lattice parameters, are openly available in the project repository. The repository also contains the trained Random Forest regressor, the Symbolic Regression pipeline scripts (based on PySR), and the complete optimization results for the candidate alloy systems. These resources can be accessed at: \url{https://github.com/mahata-lab/Ru-B2-BCC-Optimization}.

\bibliographystyle{elsarticle-num}   
\bibliography{references}           

\clearpage
\section*{Supplementary Material}
\begin{center}
\textbf{Predicting Coherent B2 Stability in Ru-Containing Refractory Alloys Through Thermodynamic--Elastic Design Maps}\\[0.6em]
Avik Mahata
\end{center}

\setcounter{table}{0}
\renewcommand{\thetable}{S\arabic{table}}
\renewcommand{\theHtable}{supp.\arabic{table}}

\setcounter{figure}{0}
\renewcommand{\thefigure}{S\arabic{figure}}
\renewcommand{\theHfigure}{supp.\arabic{figure}}

\setcounter{section}{0}
\renewcommand{\thesection}{S\arabic{section}}
\renewcommand{\theHsection}{supp.\arabic{section}}
\setcounter{equation}{0}
\renewcommand{\theequation}{S\arabic{equation}}
\renewcommand{\theHequation}{supp.\arabic{equation}}

\section{Density Functional Theory (DFT) Methodology}

\subsection{Computational Parameters and Settings}
All first-principles calculations were performed using the Vienna Ab initio Simulation Package (VASP) with the Projector Augmented Wave (PAW) method. Exchange-correlation effects were treated using the Generalized Gradient Approximation (GGA) parameterized by Perdew, Burke, and Ernzerhof (PBE).

The specific convergence criteria and lattice relaxation protocols used for the high-throughput screening are summarized in Table \ref{tab:vasp_settings}. To accurately model the chemical disorder in the multicomponent BCC and B2 phases, Special Quasirandom Structures (SQS) were generated using the \texttt{icet} package. A supercell size of $2 \times 2 \times 2$ (16 atoms) was employed for the initial screening to balance computational cost with configurational sampling, while ensuring the correlation functions matched the ideal random alloy limit for the first two nearest-neighbor shells.

\begin{table}[H]
\centering
\caption{VASP calculation parameters and SQS generation settings used in this study.}
\label{tab:vasp_settings}
\begin{tabular}{l l l}
\toprule
\textbf{Category} & \textbf{Parameter} & \textbf{Value / Setting} \\
\midrule
\multirow{4}{*}{Electronic} & Plane-wave Cutoff (\texttt{ENCUT}) & 500 eV \\
& Precision (\texttt{PREC}) & Accurate \\
& Electronic Convergence (\texttt{EDIFF}) & $10^{-6}$ eV \\
& Smearing Method (\texttt{ISMEAR}) & Methfessel-Paxton (Order 1) \\
& Smearing Width (\texttt{SIGMA}) & 0.20 eV \\
\midrule
\multirow{3}{*}{Ionic Relaxation} & Algorithm (\texttt{IBRION}) & Conjugate Gradient (2) \\
& DoF allowed (\texttt{ISIF}) & 3 (Ions + Cell Shape + Volume) \\
& Force Convergence (\texttt{EDIFFG}) & $-0.01$ eV/\AA \\
\midrule
\multirow{2}{*}{Sampling} & k-point Mesh & $6 \times 6 \times 6$ (Monkhorst-Pack) \\
& SQS Generation & \texttt{icet} (Monte Carlo) \\
\bottomrule
\end{tabular}
\end{table}

\begin{figure}[H]
    \centering
    \includegraphics[width=\textwidth]{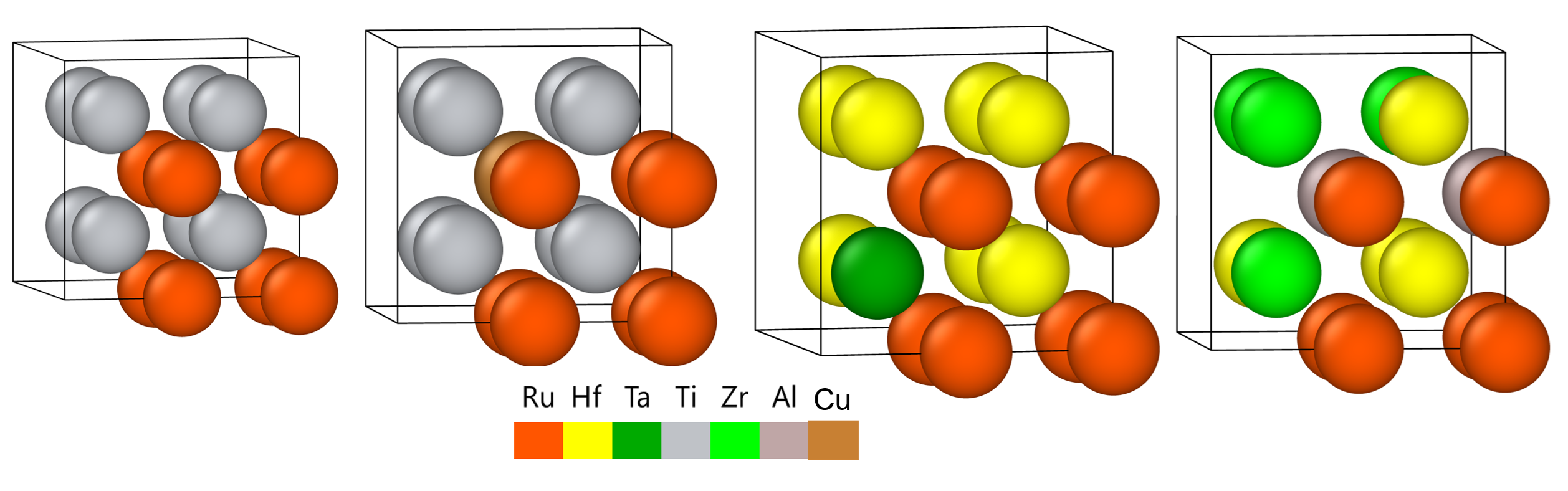}
    \caption{\rev{Real-space renderings of representative relaxed B2 supercells used in the DFT dataset. The structures illustrate the ordered Ru-based B2 binary and representative chemically substituted B2 configurations used to sample refractory-element and A-sublattice alloying. The expanded element legend identifies Ru, Hf, Ta, Ti, Zr, Al, and Cu explicitly. These atomistic renderings complement the simplified structural schematics shown as Fig.~1 in the main manuscript.}}
    \label{fig:s1_realspace}
\end{figure}

\subsection{PAW Potentials and Valence Configurations}
To ensure reproducibility of the formation energies, the specific PAW potentials utilized are listed in Table \ref{tab:paw_potentials}. Semicore $p$ and $s$ states were treated as valence electrons (using `\_pv` or `\_sv` potentials) for early transition metals to accurately capture the bonding environment in these refractory systems.

\begin{table}[H]
\centering
\caption{VASP PAW potentials and valence electron configurations.}
\label{tab:paw_potentials}
\begin{tabular}{l c c}
\toprule
\textbf{Element} & \textbf{PAW Potential Label} & \textbf{Valence Electrons ($e^-$)} \\
\midrule
Ru & Ru\_pv & $4p^6 4d^7 5s^1$ (14) \\
Nb & Nb\_pv & $4p^6 4d^4 5s^1$ (11) \\
Mo & Mo\_pv & $4p^6 4d^5 5s^1$ (12) \\
Ta & Ta\_pv & $5p^6 5d^3 6s^2$ (11) \\
W  & W\_pv  & $5p^6 5d^4 6s^2$ (12) \\
Hf & Hf\_pv & $5p^6 5d^2 6s^2$ (10) \\
Ti & Ti\_pv & $3p^6 3d^2 4s^2$ (10) \\
Zr & Zr\_sv & $4s^2 4p^6 4d^2 5s^2$ (12) \\
Al & Al     & $3s^2 3p^1$ (3) \\
V  & V\_pv  & $3p^6 3d^3 4s^2$ (11) \\
Cr & Cr\_pv & $3p^6 3d^5 4s^1$ (12) \\
\textcolor{red}{Cu} & \textcolor{red}{Cu} & \textcolor{red}{$3d^{10} 4s^1$ (11)} \\
\textcolor{red}{Si} & \textcolor{red}{Si} & \textcolor{red}{$3s^2 3p^2$ (4)} \\
\bottomrule
\end{tabular}
\end{table}


\section{Random Forest Model Development}

\subsection{Input Feature Space}
The Random Forest regressor was trained on a set of physical descriptors derived from the elemental properties and DFT-calculated parameters. Table \ref{tab:rf_features} lists the complete feature vector used for training. The inclusion of both thermodynamic (e.g., formation energy) and structural (e.g., lattice mismatch, $\delta_r$) features was critical for capturing the competition between chemical ordering and elastic strain.

\begin{table}[H]
\centering
\caption{List of input features used for the Random Forest surrogate model. Elemental properties ($\chi$, $r$) were weighted by atomic fraction on the relevant sublattices.}
\label{tab:rf_features}
\begin{tabular}{l l l}
\toprule
\textbf{Feature Class} & \textbf{Symbol} & \textbf{Description} \\
\midrule
\multirow{2}{*}{DFT Parameters} & $\Delta E_f^{B2}$ & Formation enthalpy of the B2 precipitate (eV/atom) \\
& $\delta$ & Lattice misfit between relaxed B2 and BCC phases \\
\midrule
\multirow{4}{*}{Electronic} & VEC & Valence Electron Concentration (B2 average) \\
& $\bar{\chi}$ & Average Pauling Electronegativity \\
& $\Delta \chi$ & Electronegativity difference between sublattices \\
& $\Delta H_{\mathrm{mix}}$ & Chemical mixing enthalpy (Miedema model) \\
\midrule
\multirow{2}{*}{Structural} & $\delta_r$ & Atomic size mismatch parameter ($\sqrt{\sum c_i (1 - r_i/\bar{r})^2}$) \\
& $a_{\mathrm{BCC}}$ & Lattice parameter of the matrix \\
\bottomrule
\end{tabular}
\end{table}

\subsection{Hyperparameter Optimization and Performance}
To prevent overfitting, a grid search with 5-fold cross-validation was performed. Table \ref{tab:rf_params} summarizes the search space and the optimal hyperparameters selected for the final model.

\begin{table}[H]
\centering
\caption{Hyperparameter search space and optimal settings for the Random Forest Regressor.}
\label{tab:rf_params}
\begin{tabular}{l l c}
\toprule
\textbf{Hyperparameter} & \textbf{Grid Search Range} & \textbf{Optimal Value} \\
\midrule
\texttt{n\_estimators} & [100, 200, 500, 1000] & 200 \\
\texttt{max\_depth} & [10, 15, 20, None] & 15 \\
\texttt{min\_samples\_split} & [2, 5, 10] & 5 \\
\texttt{min\_samples\_leaf} & [1, 2, 4] & 2 \\
\texttt{max\_features} & ['sqrt', 'log2', None] & 'sqrt' \\
\bottomrule
\end{tabular}
\end{table}

\subsection{Model Accuracy Metrics}
\rev{A single 80/20 hold-out split was used to evaluate both Random Forest surrogates with the hyperparameters listed in Table~\ref{tab:rf_params}. The thermodynamic surrogate was trained on the unique B2 precipitates because the formation enthalpy is a property of the precipitate alone, whereas the structural surrogate was trained on the full set of precipitate--matrix pairs because the lattice misfit depends on both phases. The resulting test-set errors are reported in Table~\ref{tab:rf_metrics}.}

\begin{table}[H]
\centering
\caption{\rev{Performance metrics for the dual-surrogate Random Forest framework using a single 80/20 hold-out split and the hyperparameters of Table~\ref{tab:rf_params}. The thermodynamic surrogate is evaluated on the 97 unique B2 precipitates and the structural surrogate on the 1746 precipitate--matrix pairs.}}
\label{tab:rf_metrics}
\begin{tabular}{l c c c}
\toprule
\textbf{Surrogate Model} & \textbf{Target Property} & \textbf{Test Set RMSE} & \textbf{Test $R^2$} \\
\midrule
$\mathcal{M}_E$ (Thermodynamics) & Formation Enthalpy ($\Delta H_f$) & $0.098$~eV/atom & $0.69$ \\
$\mathcal{M}_\delta$ (Structure) & Lattice Misfit ($\delta$) & $0.49\%$ & $0.99$ \\
\bottomrule
\end{tabular}
\end{table}

\subsubsection{Dual-Surrogate Architecture}
To independently resolve the competing drivers of stability, the machine learning framework utilizes a dual-regressor architecture rather than a single black-box model. As evidenced by the serialized models \texttt{rf\_energy.joblib} and \texttt{rf\_misfit.joblib}, two distinct Random Forest ensembles were trained in parallel:
\begin{enumerate}
    \item \textbf{Thermodynamic Surrogate ($\mathcal{M}_E$):} Predicts the formation enthalpy $\Delta H_f^{B2}$ (eV/atom).
    \item \textbf{Structural Surrogate ($\mathcal{M}_\delta$):} Predicts the lattice misfit $\delta$ (\%) relative to the BCC matrix.
\end{enumerate}
This decoupling allows for the explicit identification of "High Stability / High Misfit" candidates (structurally unviable) versus "Zero Misfit / Low Stability" candidates (thermodynamically unstable), a distinction lost in combined solvus regressors.

\subsubsection{Physics-Based Feature Importance (SHAP Analysis)}
To verify that the Random Forest model captures physical causality rather than spurious correlations, we computed SHAP (SHapley Additive exPlanations) values for the predicted lattice misfit. The top governing features identified by the model are listed in Table \ref{tab:shap_importance}.

\begin{table}[H]
\centering
\caption{Top physical descriptors driving the lattice misfit predictions, ranked by mean absolute SHAP value. The model correctly identifies atomic size mismatch ($\delta_r$) and lattice parameter difference as the dominant factors.}
\label{tab:shap_importance}
\begin{tabular}{l l l}
\toprule
\textbf{Rank} & \textbf{Feature} & \textbf{Physical Interpretation} \\
\midrule
1 & $\Delta a_{\mathrm{ideal}}$ & Ideal Vegard's law mismatch between sublattices \\
2 & $\delta_r$ (Atomic Size Mismatch) & Local lattice distortion parameter \\
3 & $\bar{r}_{B2}$ (Avg. Radius) & Average atomic radius of the precipitate \\
4 & VEC (Valence Electron Conc.) & Electronic contribution to bond length \\
5 & $\Delta \chi$ (Electronegativity) & Ionic character contribution to volume \\
\bottomrule
\end{tabular}
\end{table}

\subsubsection{Top Candidate Alloys by Design Branch}
Table \ref{tab:top_candidates} lists the most promising zero-misfit candidates identified for each primary alloy family. These compositions were selected based on minimizing the absolute lattice misfit ($|\delta| \to 0$) while maintaining a favorable thermodynamic driving force for ordering.

\begin{table}[H]
\centering
\caption{Top 5 candidate systems for Ru-Hf, Ru-Ti, and Ru-Al design branches, ranked by lattice matching. $\Delta E_f$ is the formation enthalpy of the B2 phase. Note that candidates have been strictly categorized by their primary B2 chemistry.}
\label{tab:top_candidates}
\resizebox{\textwidth}{!}{%
\begin{tabular}{llcccc}
\toprule
\textbf{Design Branch} & \textbf{BCC Matrix} & \textbf{B2 Precipitate} & \textbf{Misfit $|\delta|$ (\%)} & \textbf{$\Delta E_f$ (eV/at)} & \textbf{Notes} \\
\midrule
\multirow{5}{*}{\textbf{Ru-Hf (Heavy)}} 
& Mo$_{25}$Ta$_{75}$ & Ru$_{75}$Al$_{25}$\_Hf & 0.01 & -0.96 & Best Match \\
& Nb$_{50}$Ta$_{50}$ & Ru$_{75}$Hf$_{75}$\_Nb$_{25}$ & 0.02 & -0.99 & High Stability \\
& Mo$_{75}$Ta$_{25}$ & Ru$_{75}$Al$_{25}$\_HfTi & 0.02 & -0.90 & Complex B2 \\
& Nb$_{75}$V$_{25}$ & Ru$_{75}$Si$_{25}$\_Hf & 0.01 & -1.01 & Si-doped \\
& Nb$_{75}$Mo$_{25}$ & Ru$_{75}$Si$_{25}$\_Hf & 0.82 & -0.96 & High Entropy \\
\midrule
\multirow{5}{*}{\textbf{Ru-Ti (Light)}} 
& Ta$_{25}$V$_{75}$ & Ru$_{75}$Cu$_{25}$\_Ti & 0.03 & -0.86 & Cu-doped \\
& Nb$_{25}$V$_{75}$ & Ru$_{75}$Cu$_{25}$\_Ti & 0.07 & -0.86 & - \\
& Mo$_{50}$V$_{50}$ & Ru$_{75}$Cr$_{25}$\_Ti & 0.08 & -0.81 & Cr-doped \\
& Nb$_{25}$V$_{75}$ & Ru$_{75}$Al$_{25}$\_Ti & 0.74 & -0.94 & Al-Tuned \\
& Ta$_{25}$V$_{75}$ & Ru\_Ti$_{75}$\_V$_{25}$ & 0.62 & -0.96 & Binary-like \\
\midrule
\multirow{5}{*}{\textbf{Ru-Al (Tuned)}} 
& Ta$_{50}$V$_{50}$ & Ru$_{75}$Al$_{25}$\_HfAl & 0.06 & -1.00 & Dual-Al \\
& Nb$_{50}$V$_{50}$ & Ru$_{75}$Al$_{25}$\_HfAl & 0.18 & -1.00 & High Stability \\
& Mo$_{25}$Ta$_{75}$ & Ru$_{75}$Al$_{25}$\_Hf & 0.01 & -0.96 & Zero Misfit \\
& Nb$_{75}$Mo$_{25}$ & Ru$_{75}$Al$_{25}$\_Hf & 0.32 & -0.90 & - \\
& Nb$_{25}$V$_{75}$ & Ru$_{75}$Al$_{25}$\_Ti & 0.74 & -0.94 & Light/Tuned \\
\bottomrule
\end{tabular}%
}
\end{table}


\section{\rev{Additional Model Validation and Composition Analysis}}

\subsection{\rev{Symbolic-regression model selection}}
\rev{The symbolic-regression analysis separates the algebraic form of the fitted relation from the model-selection comparison used to identify that form. Table~\ref{tab:sr_model_selection} \revB{compares the selected PySR expression with a constrained alternative in which a quadratic misfit term is included explicitly.} The formation enthalpies and lattice misfits are supplied by the DFT feature matrix, while the symbolic-regression search and the solvus calibration target are part of the analysis pipeline referenced in the Data Availability statement.} \revB{The symbolic-regression pipeline scripts are available in that repository.}

{\color{red}
\begin{table}[H]
\centering
\caption{Symbolic-regression model-selection check. \revB{The selected linear absolute-misfit form is compared with a constrained quadratic alternative over the sampled misfit range, $|\delta|\approx0.005$--0.03 (Section~3.4 of the main manuscript).}}
\label{tab:sr_model_selection}
\revB{\begin{tabular}{l l p{0.45\textwidth}}
\toprule
Expression & Role & Outcome \\
\midrule
$A(\Delta H/\Delta S)-B|\delta|$ & selected form & retained; $B=20000$~K gives approximately 200~$^\circ$C per 1\% misfit \\
$A(\Delta H/\Delta S)-C\delta^2$ & forced quadratic comparison & $\delta^2\approx10^{-4}$--$10^{-3}$ over the sampled range, giving solvus shifts of only a few tens of $^\circ$C; not robustly identifiable on the solvus temperature scale \\
\bottomrule
\end{tabular}}
\end{table}
}

\subsection{\rev{Family-level misfit statistics}}
\rev{The family-level lattice-matching behavior was quantified directly from the Cartesian pair dataset, in which every relaxed B2 precipitate is paired with every relaxed BCC matrix. The dataset contains 100 Ru-containing B2 precipitates and 18 BCC matrices, for a total of 1800 pairs, and each family is defined by its dominant B-sublattice former. All lattice parameters are evaluated on a consistent conventional-cell basis. As an internal consistency check, the recomputed misfits of the aluminium-tuned and copper-tuned Ru--Hf candidates against the Nb$_{75}$V$_{25}$ matrix reproduce the values reported in the candidate table to within a few hundredths of one percent.}

{\color{red}
\begin{table}[H]
\centering
\caption{\rev{Distribution of absolute lattice misfit by B2 chemical family, computed from the full Cartesian dataset of 100 Ru-containing B2 precipitates paired with 18 BCC matrices. The statistics show that all four families reach the coherent window for an appreciable fraction of matrices, so no family is uniformly displaced. Ru--Hf and Ru--Zr have the largest fractions of coherent matches, while the Ru--Al family, including its aluminium-tuned members, reaches the two-percent window for 30~percent of matrix pairings.}}
\label{tab:family_misfit_stats}
\begin{tabular}{l c c c c}
\toprule
Family & Median $|\delta|$ (\%) & IQR (\%) & Fraction $|\delta|\le1\%$ & Fraction $|\delta|\le2\%$ \\
\midrule
Ru--Hf & $2.27$ & $1.00$--$4.13$ & $25\%$ & $46\%$ \\
Ru--Zr & $2.41$ & $1.17$--$5.17$ & $22\%$ & $43\%$ \\
Ru--Ti & $3.58$ & $1.71$--$5.50$ & $19\%$ & $29\%$ \\
Ru--Al & $3.45$ & $1.68$--$5.50$ & $14\%$ & $30\%$ \\
\bottomrule
\end{tabular}
\end{table}
}

\rev{The Ru--Al family has a somewhat larger median misfit than Ru--Hf and Ru--Zr, but it is not uniformly displaced from the coherent regime: 14~percent of matrix pairings place it within one percent misfit and 30~percent within two percent, which is comparable to the other families. The near-coherent Ru--Al members are the aluminium-tuned structures that pair with ruthenium-poor and vanadium-rich matrices. This family-level result is distinct from the use of aluminium as a substitutional lattice-tuning addition within an existing Ru--Hf B2 phase.}

\subsection{\rev{Al and Si A-site concentration series}}
\rev{The DFT dataset contains $(\mathrm{Ru}_{1-x}M_x)$--Hf structures at Ru fractions 1.0, 0.875, 0.75, 0.625, and 0.5 for $M=$Al and Si on a pure hafnium B sublattice. The complete substitution series is reported in Table~\ref{tab:existing_tuning_series}. The formation enthalpy of each cell includes the elemental aluminium or silicon reference as appropriate. The thermodynamic ratio is evaluated against a representative Nb$_{75}$V$_{25}$ matrix ($\Delta S_{\mathrm{mix}}=4.85\times10^{-5}$~eV/K), and the representative lattice misfit is evaluated against the same matrix.}

{\color{red}
\begin{table}[H]
\centering
\caption{\rev{A-site substitution series $(\mathrm{Ru}_{1-x}M_x)$--Hf. Lattice parameters and formation enthalpies are relaxed DFT values; the thermodynamic ratio uses the formation enthalpy of the substituted cell and the ideal configurational entropy of a representative Nb$_{75}$V$_{25}$ matrix, and the misfit is evaluated against that matrix ($a_{\mathrm{BCC}}=3.248$~\AA).}}
\label{tab:existing_tuning_series}
\begin{tabular}{l c c c c}
\toprule
Substituent & $x$ & $a_{\mathrm{B2}}$ (\AA) & $\Delta H/\Delta S$ (K) & Representative $\delta$ (\%) \\
\midrule
Al & 0.000 & $3.239$ & $24552$ & $-0.27$ \\
Al & 0.125 & $3.249$ & $21625$ & $+0.03$ \\
Al & 0.250 & $3.263$ & $19119$ & $+0.46$ \\
Al & 0.375 & $3.276$ & $17608$ & $+0.85$ \\
Al & 0.500 & $3.344$ & $14996$ & $+2.91$ \\
\midrule
Si & 0.000 & $3.239$ & $24552$ & $-0.27$ \\
Si & 0.125 & $3.243$ & $22224$ & $-0.15$ \\
Si & 0.250 & $3.307$ & $20358$ & $+1.79$ \\
Si & 0.375 & $3.304$ & $19124$ & $+1.71$ \\
Si & 0.500 & $3.341$ & $17194$ & $+2.82$ \\
\bottomrule
\end{tabular}
\end{table}
}

\rev{For the aluminium series against the Nb$_{75}$V$_{25}$ matrix, the misfit increases from slightly negative at $x=0$ through approximately zero near $x=0.125$ to $+2.9$~percent at $x=0.5$. A linear fit places the zero-misfit cross-over at $x^{*}\simeq0.11$, corresponding to approximately 11~at.\% aluminium on the A sublattice. Beyond that composition, the misfit increases while the thermodynamic ratio continues to decrease, so additional aluminium is unfavorable to the combined thermodynamic--coherency balance. The silicon series behaves similarly, with a slightly earlier cross-over near $x^{*}\simeq0.07$ and a steeper loss of lattice matching. The two series therefore show that modest substitution can improve the coherency of an initially mismatched pair, whereas excessive substitution drives the lattice beyond the matched condition while reducing the thermodynamic driving force.}

\subsection{\rev{Higher-order matrix extension}}
\rev{Transferability beyond the binary and ternary matrices is evaluated using six higher-order BCC SQS matrices while retaining the same Ru-containing B2 library. The resulting lattice parameters provide an independent matrix set for assessing pair-specific B2--BCC lattice matching at greater compositional complexity.}

{\color{red}
\begin{table}[H]
\centering
\caption{Higher-order BCC matrices used for the transferability check.}
\label{tab:higher_order_matrices}
\begin{tabular}{l c c c}
\toprule
Matrix & Target composition & Cell & $a_{\mathrm{BCC}}$ (\AA) \\
\midrule
NbMoTaW & equiatomic & 16-atom SQS & $3.2315$ \\
NbMoTaV & equiatomic & 16-atom SQS & $3.1956$ \\
NbTaVW & equiatomic & 16-atom SQS & $3.2008$ \\
MoTaVW & equiatomic & 16-atom SQS & $3.1534$ \\
NbMoVW & equiatomic & 16-atom SQS & $3.1583$ \\
NbMoTaVW & equiatomic & 20-atom SQS & $3.1887$ \\
\bottomrule
\end{tabular}
\end{table}
}

\subsection{\rev{Targeted electronic-structure calculations}}
\rev{The electronic-structure analysis focuses on five representative relaxed structures: RuHf, RuTi, RuZr, RuAl, and one Al-substituted Ru--Hf configuration. Static calculations were performed on the relaxed geometries to evaluate the projected density of states for all five structures. \revB{COHP/ICOHP and Bader analyses are not reported in the present work.}}

{\color{red}
\begin{table}[H]
\centering
\caption{Targeted static electronic-structure analyses for the five representative B2 structures.}
\label{tab:electronic_runs}
\begin{tabular}{l c c c}
\toprule
Structure & Geometry & PDOS & COHP/Bader \\
\midrule
RuHf & existing & complete & \revB{not reported} \\
RuTi & existing & complete & \revB{not reported} \\
RuZr & existing & complete & \revB{not reported} \\
RuAl & existing & complete & \revB{not reported} \\
Al-substituted Ru--Hf & existing & complete & \revB{not reported} \\
\bottomrule
\end{tabular}
\end{table}
}

\subsection{\rev{Literature benchmarking data}}
\rev{Published thermochemical, structural, and solvus measurements provide external benchmarks for the computational quantities used in the study. The literature values and their sources are summarized in Table~\ref{tab:literature_placeholders} and are used in the corresponding benchmark comparisons in the main manuscript.}

{\color{red}
\begin{table}[H]
\centering
\caption{Literature quantities used in the benchmark, with sources.}
\label{tab:literature_placeholders}
\resizebox{\textwidth}{!}{%
\begin{tabular}{l l l}
\toprule
Benchmark & Systems & Source value and reference \\
\midrule
Calorimetry & RuHf, RuTi, RuZr, RuAl & $\Delta H_f=-91.8,-77.0,-68.7,-54.5$~kJ/mol-atom \cite{GuoKleppa2001,SuNash2005} \\
Diffraction & RuHf, RuTi, RuZr, RuAl & $a=3.225, 3.060\text{--}3.070, 3.253, 2.992\text{--}3.036$~\AA \cite{Frey2024b} \\
Diffraction (BCC) & Nb, Mo, Ta, W, V & $a=3.301, 3.147, 3.303, 3.165, 3.026$~\AA \cite{Frey2024b} \\
Pseudo-binary tuning & Ru--X B2 database & composition vs lattice parameter \cite{Kube2024} \\
Solvus comparison & Hf--Ru, Zr--Ru, Ti--Ru BCC--B2 alloys & measured ranges and misfits \cite{Frey2024b,Kube2024} \\
Dilute alloy & Hf$_3$Ru$_2$--Nb$_{67}$V$_{28}$ & no B2 down to 1000~$^\circ$C (terminal solubility) \cite{Frey2024b} \\
\bottomrule
\end{tabular}%
}
\end{table}
}

\section{\rev{Additional First-Principles Results}}

\rev{This section summarizes the matched elemental reference-state calculations, higher-order BCC matrix calculations, and targeted electronic-structure calculations used in the analysis. All reported values are taken directly from normally terminated, force-converged VASP runs. Absolute energies of the higher-order matrices are reported as reproducible calculation outputs rather than as formation energies, because a composition-matched elemental reference set is required before energies of different compositions can be compared as formation enthalpies.}

\subsection{\rev{Elemental reference states}}
\rev{Matched-settings static calculations were performed for the hexagonal close-packed ground state and the body-centred-cubic reference of Ru, Hf, Ti, and Zr at the production cutoff. The resulting energies are listed in the reference-state table of the main manuscript. For each element, the body-centred-cubic energy lies above the hexagonal close-packed energy, confirming the HCP ground state within the present calculations and quantifying the reference shift used to place the RuHf, RuTi, and RuZr formation enthalpies on an elemental ground-state scale. The B2 formation enthalpy on the body-centred-cubic reference is defined for a compound $\mathrm{RuX}$ as $\Delta H_{\mathrm{f}}^{\mathrm{BCC}} = E_{\mathrm{RuX}} - \tfrac{1}{2}E_{\mathrm{Ru}}^{\mathrm{BCC}} - \tfrac{1}{2}E_{\mathrm{X}}^{\mathrm{BCC}}$, and the ground-state-referenced value replaces each body-centred-cubic reference energy with the hexagonal close-packed energy of the same element.}

\subsection{\rev{Higher-order BCC matrices}}
\rev{Five equiatomic quaternary special quasirandom matrices and the equiatomic quinary NbMoTaVW matrix were relaxed and evaluated using the same first-principles workflow. Their relaxed BCC-equivalent lattice parameters and static energies are listed in Table~\ref{tab:higher_order_completed}. All six relaxations reached the requested force criterion with small residual pressures and small volume changes, and both the relaxation and static calculations terminated normally.}

{\color{red}
\begin{table}[H]
\centering
\caption{Higher-order BCC SQS matrices. The static energy per atom is a reproducible calculation output; it is not a formation energy. The BCC-equivalent lattice parameter is obtained from the relaxed cell volume per atom.}
\label{tab:higher_order_completed}
\begin{tabular}{l l c c c c}
\toprule
ID & Composition & Atoms & Static energy (eV/atom) & $a_{\mathrm{BCC}}$ (\AA) & QA \\
\midrule
B01 & NbMoTaW & 16 & $-11.5565$ & $3.2315$ & passed \\
B02 & NbMoTaV & 16 & $-10.4807$ & $3.1956$ & passed \\
B03 & NbTaVW & 16 & $-10.9652$ & $3.2008$ & passed \\
B04 & MoTaVW & 16 & $-11.2615$ & $3.1534$ & passed \\
B05 & NbMoVW & 16 & $-10.7875$ & $3.1583$ & passed \\
B06 & NbMoTaVW & 20 & $-11.0120$ & $3.1887$ & passed \\
\bottomrule
\end{tabular}
\end{table}
}

\subsection{\rev{Targeted electronic-structure analysis}}
\rev{Static electronic-structure calculations with tetrahedron integration and projected orbital character were performed for the five representative structures. Table~\ref{tab:electronic_completed} summarizes the Ru $d$-band centre relative to the Fermi level and the total density of states at the Fermi level per atom, together with the relaxed lattice parameter and, for the three early-transition-metal binaries, the formation enthalpy on both reference scales. RuHf, RuTi, and RuZr share a pronounced pseudogap and relatively low $N(E_{\mathrm{F}})$, whereas RuAl shows a markedly higher density of states at the Fermi level and additional Ru--Al hybridization. The Al-substituted Ru--Hf structure has an intermediate $N(E_{\mathrm{F}})$, while its Ru $d$-band centre is deeper than those of the four binary compounds. These results show that the electronic descriptors do not vary monotonically with Al substitution and provide supporting interpretation for the accompanying structural and thermodynamic trends.}

{\color{red}
\begin{table}[H]
\centering
\caption{Summary of the electronic-structure calculations. \revB{Ground-state-referenced formation enthalpies use HCP Ru, Hf, Ti, and Zr and FCC Al; the RuAl value corresponds to the calorimetric benchmark in the main manuscript ($-62.2$~kJ/mol-atom). The ground-state-referenced value was not evaluated for the Al-substituted cell.}}
\label{tab:electronic_completed}
\resizebox{\textwidth}{!}{%
\begin{tabular}{l c c c c c}
\toprule
Structure & Ru $d$-band centre (eV) & $N(E_{\mathrm{F}})$ (eV$^{-1}$ atom$^{-1}$) & $\Delta H_{\mathrm{f}}^{\mathrm{BCC}}$ (eV/atom) & $\Delta H_{\mathrm{f}}^{\mathrm{gs}}$ (eV/atom) & $a$ (\AA) \\
\midrule
RuHf & $-1.32$ & $0.22$ & $-1.190$ & $-0.842$ & $3.239$ \\
RuTi & $-1.42$ & $0.26$ & $-1.053$ & $-0.758$ & $3.074$ \\
RuZr & $-1.26$ & $0.30$ & $-0.965$ & $-0.666$ & $3.269$ \\
RuAl & $-1.06$ & $0.86$ & $-1.009$ & \revB{$-0.645$} & $3.009$ \\
(Ru$_{0.75}$Al$_{0.25}$)--Hf & $-1.48$ & $0.58$ & $-0.926$ & \revB{--} & $3.265$ \\
\bottomrule
\end{tabular}%
}
\end{table}
}

\section{Symbolic Regression Arithmetic Verification}

To demonstrate the arithmetic used to obtain the effective solvus temperatures presented in the main manuscript, explicit calculations are provided for three representative candidates using the derived relation. \rev{In this equation $\Delta H$ is used as a positive magnitude; the negative numbers listed in the bullets below are the raw DFT formation energies. These worked examples illustrate internal candidate calculations and are distinct from the chemistry-matched literature proxies used for the experimental benchmark.}
\begin{equation}
    T_{\mathrm{solvus}} (^{\circ}\mathrm{C}) \approx 0.1121 \left( \frac{\Delta H_{\mathrm{form}}}{\Delta S_{\mathrm{mix}}} \right) - 20000 \cdot |\delta| - 273.15
\end{equation}
Values for $\Delta H$, $\Delta S$, and misfit $|\delta|$ are taken directly from the dataset (Table \ref{tab:full_candidates}).

\subsection*{Example 1: High Stability (Lattice Tuned Ru-Hf)}
\textbf{System:} Nb$_{75}$V$_{25}$ Matrix + Ru$_{75}$Cu$_{25}$Hf Precipitate.
\begin{itemize}
    \item \textbf{Thermodynamics:} $\Delta H = -0.969$ eV/atom; $\Delta S = 4.85 \times 10^{-5}$ eV/K.
    \item \textbf{Ratio ($\Delta H / \Delta S$):} $\mathbf{20,004}$ K.
    \item \textbf{Misfit ($|\delta|$):} $\mathbf{0.00367}$ (0.36\%).
\end{itemize}
\textbf{Calculation:}
$$ T \approx 0.1121(20004) - 20000(0.00367) - 273.15 $$
$$ T \approx 2242.4 - 73.4 - 273.15 = \mathbf{1896 \ ^{\circ}\mathrm{C}} $$
\textit{(Representative high-stability candidate from the screening dataset.)}

\subsection*{Example 2: Intermediate Stability (Ru-Zr)}
\textbf{System:} Nb$_{75}$V$_{25}$ Matrix + Ru$_{75}$Cu$_{25}$Zr Precipitate.
\begin{itemize}
    \item \textbf{Thermodynamics:} $\Delta H = -0.791$ eV/atom; $\Delta S = 4.85 \times 10^{-5}$ eV/K.
    \item \textbf{Ratio ($\Delta H / \Delta S$):} $\mathbf{16,321}$ K.
    \item \textbf{Misfit ($|\delta|$):} $\mathbf{0.00764}$ (0.76\%).
\end{itemize}
\textbf{Calculation:}
$$ T \approx 0.1121(16321) - 20000(0.00764) - 273.15 $$
$$ T \approx 1829.6 - 152.8 - 273.15 = \mathbf{1404 \ ^{\circ}\mathrm{C}} $$
\textit{(Representative intermediate-stability Ru--Zr candidate.)}

\subsection*{Example 3: Destabilized System (Cr-Doped)}
\textbf{System:} Nb$_{75}$V$_{25}$ Matrix + Ru$_{75}$Cr$_{25}$Zr Precipitate.
\begin{itemize}
    \item \textbf{Thermodynamics:} $\Delta H = -0.721$ eV/atom; $\Delta S = 4.85 \times 10^{-5}$ eV/K.
    \item \textbf{Ratio ($\Delta H / \Delta S$):} $\mathbf{14,897}$ K (Significantly reduced).
    \item \textbf{Misfit ($|\delta|$):} $\mathbf{0.0042}$ (0.42\%).
\end{itemize}
\textbf{Calculation:}
$$ T \approx 0.1121(14897) - 20000(0.0042) - 273.15 $$
$$ T \approx 1669.9 - 84.0 - 273.15 = \mathbf{1313 \ ^{\circ}\mathrm{C}} $$
\textit{(Demonstrates the solvus suppression in Cr-containing systems)}

\clearpage
\section{Extended Dataset of Ideal Candidates}
Table \ref{tab:full_candidates} provides the crystallographic and energetic data for the top candidates identified in this study.

\begin{small}
\begin{longtable}{llcccc}
\caption{Extended list of B2-BCC candidates. $\Delta E_f$ values are in eV/atom. Misfit is the absolute linear mismatch $|\delta|$. $T_{\mathrm{solv}}$ is the SR-predicted solvus temperature.} \label{tab:full_candidates} \\
\toprule
\textbf{BCC Matrix} & \textbf{B2 Precipitate} & \textbf{Misfit $|\delta|$} & \textbf{$\Delta E_f^{B2}$} & \textbf{$\Delta E_f^{BCC}$} & \textbf{$T_{\mathrm{solv}}$ ($^{\circ}$C)} \\
\midrule
\endfirsthead
\toprule
\textbf{BCC Matrix} & \textbf{B2 Precipitate} & \textbf{Misfit $|\delta|$} & \textbf{$\Delta E_f^{B2}$} & \textbf{$\Delta E_f^{BCC}$} & \textbf{$T_{\mathrm{solv}}$ ($^{\circ}$C)} \\
\midrule
\endhead
\bottomrule
\endfoot

Nb75-V25 & Ru75Cu25\_Hf & 0.0037 & -0.97 & 0.05 & 1897 \\
Ta25-V75 & Ru\_Ti75\_V25 & 0.0062 & -0.96 & 0.05 & 1825 \\
Ta75-V25 & Ru\_Zr75\_Ta25 & 0.0020 & -0.92 & 0.06 & 1819 \\
Ta75-V25 & Ru75Cr25\_Hf & 0.0039 & -0.94 & 0.06 & 1814 \\
Mo25-Ta75 & Ru75Si25\_Hf & 0.0049 & -0.94 & -0.07 & 1812 \\
Nb25-V75 & Ru\_Ti75\_V25 & 0.0072 & -0.96 & 0.06 & 1808 \\
Nb75-V25 & Ru\_Zr75\_Ta25 & 0.0011 & -0.91 & 0.05 & 1808 \\
Nb25-Mo75 & Ru\_Hf75\_Ta25 & 0.0080 & -0.96 & -0.07 & 1794 \\
Mo25-Ta75 & Ru75Al25\_Hf & 0.0001 & -0.89 & -0.07 & 1782 \\
Nb75-Mo25 & Ru75Si25\_Hf & 0.0082 & -0.96 & -0.05 & 1779 \\
Nb25-V75 & Ru\_Ti75\_Nb25 & 0.0084 & -0.96 & 0.06 & 1779 \\
Nb25-V75 & Ru75Al25\_Ti & 0.0074 & -0.94 & 0.06 & 1765 \\
Ta25-V75 & Ru\_Ti75\_Nb25 & 0.0094 & -0.96 & 0.05 & 1756 \\
Nb75-Mo25 & Ru75Al25\_Hf & 0.0032 & -0.90 & -0.05 & 1755 \\
Ta25-V75 & Ru75Al25\_Ti & 0.0084 & -0.94 & 0.05 & 1742 \\
Nb75-V25 & Ru75Cr25\_Hf & 0.0070 & -0.92 & 0.05 & 1725 \\
Nb75-V25 & Ru\_Zr75\_Nb25 & 0.0002 & -0.86 & 0.05 & 1723 \\
Ta25-V75 & Ru75Cu25\_Ti & 0.0003 & -0.86 & 0.05 & 1704 \\
Nb25-V75 & Ru75Cu25\_Ti & 0.0007 & -0.86 & 0.06 & 1697 \\
Nb25-Ta75 & Ru75Si25\_Hf & 0.0193 & -1.02 & 0.01 & 1696 \\
Ta75-V25 & Ru\_Hf75\_V25 & 0.0179 & -1.00 & 0.06 & 1694 \\
Nb25-Mo75 & Ru\_Hf75\_V25 & 0.0032 & -0.88 & -0.07 & 1690 \\
Ta75-V25 & Ru\_Zr75\_Nb25 & 0.0032 & -0.88 & 0.06 & 1689 \\
Mo75-V25 & Ru\_Ti75\_Ta25 & 0.0030 & -0.87 & -0.09 & 1677 \\
Mo75-Ta25 & Ru\_Hf75\_Ta25 & 0.0096 & -0.93 & -0.11 & 1676 \\
Nb25-Ta75 & Ru75Al25\_Hf & 0.0143 & -0.96 & 0.01 & 1671 \\
Nb25-Mo75 & Ru\_Hf75\_Nb25 & 0.0094 & -0.92 & -0.07 & 1668 \\
Mo25-Ta75 & Ru\_Hf75\_Ta25 & 0.0148 & -0.97 & -0.07 & 1667 \\
Nb25-V75 & Ru\_Ti75\_W25 & 0.0017 & -0.85 & 0.06 & 1650 \\
Ta25-V75 & Ru75Cr25\_Ti & 0.0040 & -0.86 & 0.05 & 1646 \\
Nb75-V25 & Ru75\_Al25\_HfZr & 0.0109 & -0.92 & 0.05 & 1642 \\
Mo75-Ta25 & Ru\_Hf75\_V25 & 0.0016 & -0.84 & -0.11 & 1638 \\
Nb75-Mo25 & Ru\_Hf75\_Ta25 & 0.0181 & -0.98 & -0.05 & 1635 \\
Nb25-V75 & Ru75Cr25\_Ti & 0.0051 & -0.87 & 0.06 & 1629 \\
Ta25-V75 & Ru\_Ti75\_W25 & 0.0027 & -0.84 & 0.05 & 1627 \\
Ta75-V25 & Ru75\_Al25\_HfTi & 0.0161 & -0.96 & 0.06 & 1621 \\
Nb25-Mo75 & Ru75\_Al25\_HfTi & 0.0015 & -0.83 & -0.07 & 1617 \\
Ta75-V25 & Ru75\_Al25\_HfZr & 0.0140 & -0.93 & 0.06 & 1607 \\
Nb25-Ta75 & Ru75\_Al25\_HfZr & 0.0085 & -0.88 & 0.01 & 1603 \\
Ta50-V50 & Ru75\_Al25\_HfAl & 0.0006 & -1.00 & 0.07 & 1602 \\
Mo25-Ta75 & Ru\_Hf75\_Nb25 & 0.0133 & -0.92 & -0.07 & 1599 \\
Nb75-Ta25 & Ru75Al25\_Hf & 0.0165 & -0.95 & -0.01 & 1597 \\
Nb75-V25 & Ru75Si25\_Zr & 0.0114 & -0.90 & 0.05 & 1589 \\
Nb75-Mo25 & Ru75\_Al25\_HfZr & 0.0026 & -0.82 & -0.05 & 1582 \\
Nb25-V75 & Ru\_Ti75\_Mo25 & 0.0022 & -0.82 & 0.06 & 1577 \\
Mo75-V25 & Ru\_Ti75\_Nb25 & 0.0017 & -0.81 & -0.09 & 1577 \\
Nb25-Ta75 & Ru75Si25\_Zr & 0.0080 & -0.87 & 0.01 & 1569 \\
Nb50-V50 & Ru75\_Al25\_HfAl & 0.0018 & -1.00 & 0.07 & 1568 \\
Nb75-Mo25 & Ru\_Hf75\_Nb25 & 0.0166 & -0.94 & -0.05 & 1566 \\
Ta75-V25 & Ru\_Hf75\_W25 & 0.0116 & -0.89 & 0.06 & 1565 \\
Nb50-V50 & Ru\_Hf75\_Ta25 & 0.0157 & -1.10 & 0.07 & 1476 \\

\end{longtable}
\end{small}

\end{document}